\documentclass[conference]{IEEEtran}

\usepackage{comment}
\usepackage{cite}
\usepackage{amsmath,amssymb,amsfonts}
\usepackage{graphicx,color}
\usepackage{textcomp}
\usepackage[colorlinks=true,linkcolor=blue]{hyperref}
\usepackage{url}
\usepackage[ruled,vlined,linesnumbered]{algorithm2e}
\usepackage{xcolor}
\usepackage{fancyhdr}
\fancypagestyle{firstpage}{
    \fancyhf{}
    
  \fancyhead[C]{
    \fbox{
        \parbox{0.9\textwidth}{
            \centering
            \small
            This manuscript has been submitted for peer review and possible publication in an IEEE journal.\\[1pt]
            The content herein represents the version prepared by the authors and may be subject to further revision during the review.
        }
    }
}
}

\begin{document}

\title{ReVeal-MT: A Physics-Informed Neural Network for Multi-Transmitter Radio Environment Mapping}

\author{Mukaram Shahid, 
        Kunal Das, 
        Hadia Ushaq, 
        Hongwei Zhang, 
        Jiming Song, 
        Daji Qiao,\\ 
        Sarath Babu, 
        Yong Guan, 
        Zhengyuan Zhu, and 
        Arsalan Ahmad}

\maketitle

\thispagestyle{firstpage}

\begin{abstract}
Accurately mapping the radio environment (e.g., identifying wireless signal strength at specific frequency bands and geographic locations) is crucial for efficient spectrum sharing, enabling Secondary Users~(SUs) to access underutilized spectrum bands while protecting Primary Users~(PUs). While existing models have made progress, they often degrade in performance when multiple transmitters coexist, due to the compounded effects of shadowing, interference  from adjacent transmitters. To address this challenge, we extend our prior work on Physics-Informed Neural Networks~(PINNs) for single-transmitter mapping to derive a new multi-transmitter Partial Differential Equation~(PDE) formulation of the Received Signal Strength Indicator~(RSSI). We then propose \emph{ReVeal-MT} (Re-constructor and Visualizer of Spectrum Landscape for Multiple Transmitters), a novel PINN which integrates the multi-source PDE residual into a neural network loss function, enabling accurate spectrum landscape reconstruction from sparse RF sensor measurements. ReVeal-MT is validated using real-world measurements from the ARA wireless living lab across rural and suburban environments, and benchmarked against 3GPP and ITU-R channel models and a baseline PINN model for a  single transmitter use-case. Results show that ReVeal-MT achieves substantial accuracy gains in multi-transmitter scenarios, e.g., achieving an RMSE of only 2.66\,dB with as few as 45 samples over a 370-square-kilometer region, while maintaining low computational complexity. These findings demonstrate that ReVeal-MT significantly advances radio environment mapping under realistic multi-transmitter conditions, with strong potential for enabling fine-grained spectrum management and precise coexistence between PUs and SUs.

\end{abstract}

\section{INTRODUCTION}

\IEEEPARstart{E}{xisting} spectrum sharing frameworks, such as those implemented in the TV White Space (TVWS) database and  Citizens Broadband Radio Service (CBRS) Spectrum Access System (SAS), rely heavily on traditional statistical models. However, such models struggle to accurately capture the real-world spectrum occupancy and do not generalize well enough to capture shadowing and fading caused by different kinds of terrain and environmental conditions, leading to conservative approaches that over-protect the primary users~(PUs) and cause discrepancies in channel availability for spectrum re-use \cite{shahid2024,DB_Critique,DB_Implementation}. 
    In the meantime, deterministic models such as ray tracing 
require precise characterization of the complete propagation environment such as vegetation, trees, buildings,  and material properties. Any errors in accurately defining these site-specific characteristics can degrade the models' accuracy. In addition, such deterministic models are computationally expensive to be useful for at-scale, online spectrum management in dynamic radio environments. 
    The existing stochastic and deterministic models also typically require the transmitter's operational parameters, such as Effective Isotropic Radiated Power~(EIRP), transmitter location, and antenna characteristics, which may not be available in real-world scenarios (e.g., where strong privacy or military secrecy are desired). 
The aforementioned drawbacks call for new models that are generically applicable to diverse environments and that are highly accurate in capturing the impact of transmitters and environmental factors (e.g., vegetation, trees, and buildings) on receiver signal strength while not requiring comprehensive, highly accurate information about the transmitters and environment.

To address the above challenge, data-driven modeling via Spectrum Cartography (SC) offers a promising solution avenue. 
In SC, ground-truth wireless signal measurements from sparsely distributed RF  sensors are used to accurately generate the Radio Environment Map~(REM) in the geographical area of interest~\cite{shahid2024,Bhattarai2018,fcc2020,Spectrum_carography,cartography_techniques}. 
    In particular, SC treats radio environment mapping as an ill-posed inverse problem where transmitter location and RF parameters are not available, and SC uses the spatial relationship between measurements to regenerate REMs~\cite{Subash,SC_survey,NTIA_spectrum_cartography}. The generated REMs have a wide range of applications in wireless communications, for instance, dynamically identifying white spaces for efficient spectrum sharing, optimizing power control for interference management, and facilitating seamless handover~\cite{cartography_techniques,Spectrum_carography}.

However, a critical limitation of most existing SC techniques, including our previous work \cite{ReVeal}, is their focus on environments with a single dominant transmitter. In real-world wireless networks, multiple transmitters often operate non-coherently within the same frequency band and geographic area. For instance, in a rural community leveraging TV White Spaces (TVWS) for broadband access, a single area may contain multiple secondary user (SU) transmitters: fixed wireless access points providing internet to farms, sensors for agricultural IoT monitoring soil moisture and crop health, and equipment for precision irrigation systems \cite{GNN_Multi_Source}. The signals from these diverse SUs, all operating in the same underutilized band, create a complex and overlapping radio landscape. This superposition results in intricate interference patterns and co-channel coverage that cannot be captured by models designed for a single, isolated transmitter. Accurately characterizing this multi-transmitter environment is therefore essential for reliable spectrum sharing, preventing harmful interference between SUs themselves, and is a critical prerequisite for the practical, fine-grained management of spectrum resources in next-generation rural wireless networks.

Despite their promises, existing methods for generating REMs have significant limitations. For instance, techniques such as kriging and tensor decomposition assume a uniform spatial structure, failing to capture complex variations in signal strength often observed in real-world scenarios. In addition, these models typically require dense data, leading to high  %increasing 
computational and sensing costs \cite{Subash}, \cite{Block_tensor_decomposition}. Similarly, while deep learning approaches are powerful for matrix or tensor completion tasks, they often lack interpretability and require vast amounts of labeled data, which are impractical to collect in real-world settings.

To fill the gap in radio environment mapping, we propose \emph{ReVeal-MT}, an extended Physics-Informed Neural Network (PINN) architecture for blind spectrum cartography in the presence of multiple transmitters. ReVeal-MT builds on our earlier single-transmitter work by deriving a new PDE formulation for received signal strength under multiple transmitters and incorporating this PDE as a physics-based constraint into the learning framework. This innovation enables ReVeal-MT to model the spatial superposition of signals while capturing real-world variations caused by shadowing and other propagation phenomena without requiring prior knowledge of transmitter parameters or detailed environmental information (e.g., terrain). The key contributions of this paper are as follows:
\begin{itemize}
    \item \textbf{Novel Multi-Transmitter PDE Formulation:} Unlike prior PINN applications in wireless communications that rely on single-source formulations or require transmitter parameters, we derive a new second-order PDE that explicitly models the spatial superposition and interference patterns from multiple unknown transmitters. This formulation captures the Laplacian behavior of aggregate received power distributions, enabling principled handling of co-channel interference without prior knowledge of transmitter characteristics.

    \item\textbf{Spectrum Cartography with Limited Prior Information:} While existing PINN approaches for channel modeling typically require transmitter parameters (location, power, antenna patterns), ReVeal-MT operates with minimal prior knowledge of transmitter characteristics. Our method jointly learns both the radio environment map and the unknown transmitter parameters during training, making it suitable for practical scenarios where complete transmitter information is unavailable due to privacy or dynamic spectrum access constraints.
    
    \item \textbf{Data- and Computation-Efficient Mapping:} ReVeal-MT requires significantly fewer sample points than traditional techniques and achieves high accuracy. Its data efficiency, fast convergence, and optimized architecture make it well-suited for large-scale, real-time spectrum management in dynamic and dense radio environments.
    \item \textbf{Real-World Outdoor Evaluation:} We validate ReVeal-MT on real-world data from the ARA testbed~\cite{ARA_design_implementation} across diverse rural and suburban terrains and channel conditions. We benchmark its performance against statistical, deterministic, geospatial, and neural network models, showing that ReVeal-MT achieves substantial accuracy gains in multi-transmitter settings.
\end{itemize}

The rest of the paper is structured as follows:  Section~\ref{sec:related_work} discusses %of the article touches on the 
related work, % and prior art in the fields of channel modeling and spectrum cartography, and PINNs. 
Section~\ref{sec:methodology} introduces ReVeal, %will touch on the methodology and network architecture. 
Section~\ref{sec:experimental_setup} outlines the experiment evaluation plan, %and will touch on the Experimental Setup and Evaluation of the model under different scenarios. 
Section~\ref{sec:results} presents the experimental results, %will discuss the Results collected from the evaluation of the proposed model, 
and Section~\ref{sec:conclusion} concludes the paper. % will summarize the concluding remarks of the paper.  %\hz{do not hard code references for section, figure, table IDs; use \\label \\ref etc}

\section{Related Work }% and Prior Art in Existing Modeling Techniques}

\label{sec:related_work}

In what follows, we first review the stochastic and deterministic models typically used in today's spectrum management practice. Furthermore, we discuss geospatial models, deep learning models, and physics-informed deep learning approaches.

\paragraph{Stochastic and Deterministic Models}

Channel modeling has been a fundamental component of wireless communication systems design, providing critical insights into signal propagation for interference management and network optimization~\cite{Channel_measurement_survey}. Stochastic modeling techniques rely on  statistical distributions or empirical equations to characterize signal distribution based on the operational parameters of the transmitter (e.g., location, height, and antenna azimuth), and line-of-sight information between the transmitter and the receiver. 
The use of probabilistic techniques and statistical distributions in these models helps reduce  computational complexity.
However, such a simplification often comes at the expense of accuracy~\cite{Tataria2020}. Since these models rely on summary statistical distributions rather than specific real-world instances, they fail to capture site-specific environmental features (e.g., vegetation, trees, and buildings). As a result, these models cannot accurately characterize wireless channel behavior at a given location, where signal propagation and shadowing effects can vary significantly across space. 

Stochastic path-loss models such as the Okumura or Hata-Davidson models are developed from extensive data-driven measurement studies, where regression analysis is applied to fit functions that best capture observed behavior. While these models reduce computational complexity, they often fail to account for the diverse site-specific environmental factors that significantly impact real-world signal coverage. Similarly, deterministic models such as ray tracing can, in principle, capture wave behavior at the receiver by incorporating multipath effects including reflection, diffraction, and scattering. However, these methods demand precise site characterization (e.g., vegetation, building materials) and are computationally intensive, limiting their feasibility for at-scale, online spectrum management or for spectrum sharing at fine-grained time scales. Moreover, both stochastic and deterministic models typically require transmitter-side parameters such as location, antenna pattern, and EIRP, which may not be available in Radio Dynamic Zones (RDZs). Most importantly, these models lack the integration of real-world RF sensing data, which could otherwise provide a holistic, data-informed view of spectrum occupancy.

Deterministic models, such as ray tracing,  are site-specific and enable  precise modeling of the propagation environment by considering the geographical scene, material properties, and the scattering  between the transmitter and receiver~\cite{ray_tracing_based_model,Nueral_Ray_Tracing}. By incorporating the principles of physics such as reflection, diffraction, and scattering, deterministic models can accurately calculate the path loss, delay, and angle of each reflected component reaching the receiver. However, precision comes at the expense of computational complexity. Moreover, in real-world scenarios where it is difficult to precisely characterize the environment,  deterministic models may still result in significant, as we will demonstrate in Section~\ref{sec:results}. 

Both stochastic and deterministic models require prior operational information about the transmitter, such as height, azimuth, and EIRP, which may not be realistic in Radio Dynamic Zones~(RDZs) where multiple users utilize spectrum as a shared common resource pool. Moreover, such  models are unable to integrate real-time spectrum usage data from %fixed 
RF sensors deployed in RDZs~\cite{RDZ}.

\paragraph{Geospatial %, Machine Learning, 
and Deep Learning Models}

Geospatial interpolation techniques have been the center of attention among the wireless community for generating REMs. Approaches such as kriging~\cite{Kriging} and inverse weighted distance have been of use in modeling the spectrum occupancy based on the sparsely collected data points. However, such techniques work on the assumption of spatial stationarity and struggle to capture the nonlinear relationships, often experienced in modeling wireless channels, due to the presence of shadowing and wireless interference~\cite{Kriging}. Furthermore, these models generally lack the ability to accommodate %the model to the
new terrains and varying spatial resolutions. 

Statistical interpolation techniques, such as kriging, have been employed to generate radio environment maps by estimating unknown values from sparsely collected samples. Kriging is based on a weighted average of known points, assuming a stationary underlying process. While this method has been explored for spectrum occupancy modeling, its accuracy diminishes in environments with irregular terrain, multi-network interference, or nonstationary channel behavior. Similarly, tensor decomposition approaches, where multi-band spectrum data is represented as a three-dimensional tensor (space, time, frequency), can identify latent structures and interpolate missing entries. Yet these methods also rely on stationarity assumptions and incur high computational costs, making them unsuitable for rapid REM generation.

Deep Learning~(DL) algorithms, on the other hand, are able to learn complex nonlinear spatial relationships from sparse training data~\cite{DeepREM,ProSpire}. Various DL models such as Convolutional Neural Network~\cite{CNN}, U-Net~\cite{U_NET}, and Generative Adversarial Network~\cite{GAN} have been proposed to generate spatio-temporal spectrum maps. However, DL approaches require significant amount of training data, and collecting data from real-world deployments is a time-consuming task. Furthermore, achieving a fixed dense deployment of RF sensors is often impractical due to cost and data collection overhead.

Moreover, deep learning methods tend to be site- and frequency-agnostic, limiting their ability to generalize across diverse geographical and spectral settings. To mitigate the lack of real-world training data, many studies rely on simulated datasets derived from stochastic models. However, such datasets fail to capture the true spatiotemporal dependencies of the wireless environment. Handling sparsity across large RDZs while maintaining accuracy therefore remains a critical challenge.

\paragraph{Physics-Informed Deep Learning}

Physics-Informed Deep Learning~(PIDL) has emerged as a new compelling method to solve PDEs for both forward and inverse problems. Finite Element Methods~(FEMs) have been the key in solving PDEs in different engineering problems. However, while solving PDEs, FEMs are not capable of integrating real-world data without complex computationally expensive data assimilation techniques \cite{PINN_to_PIKAN}. Such a limitation prevents FEMs from fully utilizing measurement data, which can cause valuable system insights to be overlooked \cite{Possion_FEM}. 
    In contrast, neural networks are naturally suited for data assimilation, as they can be trained using data of varying fidelity and modality. PINNs have been developed to bridge the gap between data-driven and physics-based methods, especially in cases where partial knowledge of the physical laws and sparse measurement data are available. By embedding physical laws directly into the neural network through %an added 
    residual loss terms in the objective function, PINNs can enforce the governing PDEs as soft constraints, which enable PINNs to solve forward and inverse problems using sparse and noisy data \cite{PINN_RAISSI}. 

PINNs so far have been widely used in applications ranging from acoustic engineering to the modeling of flow dynamics and modeling of electromagnetic fields ~\cite{PINN_poisson}, however, their application in wireless channel modeling remains relatively limited.

In recent works, a number of studies have demonstrated the potential of PINNs in wireless channel modeling. Authors in \cite{PINN_wireless_1} introduced a generative PINN framework that incorporates electromagnetic constraints into the modeling of wireless channels, demonstrating strong generalization performance under sparse data conditions. Jiang et al. \cite{PINN_Wireless_2} used PINNs to solve electromagnetic integral equations for accurate path loss estimation. Additionally, Kumar and Supriya \cite{PINN_wireless_3} proposed a physics-guided deep reinforcement learning model to estimate attenuation while preserving physical constraints.

Unlike prior PINN-based approaches \cite{OJAP_PINN}, \cite{PINN_maths} which typically assume complete knowledge of boundary conditions or transmitter parameters, ReVeal-MT is designed to operate in a fully blind setting with multiple unknown transmitters. To our knowledge, this represents the first application of PINNs to multi-transmitter spectrum cartography, where the transmitter locations and power levels are jointly inferred alongside the radio environment map. In this work, we propose a novel PINN-based architecture for spectrum cartography in the C-band, leveraging real-world RF sensing data collected from the ARA testbed. Our approach demonstrates how physics-informed learning can enable accurate and scalable spectrum mapping under practical deployment constraints, including sparse sensing and limited a-priori knowledge.

\section{ReVeal-MT: PINN for Multi-Transmitter Radio Environment Mapping}
\label{sec:methodology}

Our prior work, ReVeal \cite{ReVeal}, pioneered a blind spectrum cartography method using Physics-Informed Neural Networks (PINNs) for single-transmitter scenarios. By formulating a governing PDE from a path-loss model, ReVeal accurately reconstructed radio maps from sparse measurements without transmitter or environmental priors, significantly outperforming existing approaches.

While these results validated the efficacy of ReVeal for single-transmitter scenarios, real-world wireless networks often involve multiple transmitters operating non-coherently within overlapping coverage regions. In such settings, receiver signal strength is no longer governed by a single source but by the aggregate contributions of multiple transmitters and the resulting interference patterns. These overlapping transmissions create spatially heterogeneous interference zones that cannot be accurately modeled by extending single-source PDE formulations. Therefore, advancing spectrum cartography to multi-transmitter environments is essential for capturing realistic wireless conditions and supporting applications such as interference management, dynamic spectrum sharing, and seamless handover in dense deployments.

To address this gap, we extend ReVeal to incorporate multi-transmitter system modeling. In this new formulation, each transmitter contributes independently to the received signal, and the aggregate field is modeled as a combination of these contributions. The derived PDE now includes additional terms that capture cross-interference and spatial gradient interactions between transmitters. This generalization preserves the physics-informed foundation of ReVeal while enabling it to scale to complex, real-world environments where multiple transmitters coexist in the same frequency band.

\begin{figure}[!htbp]
        \centering
        \includegraphics[width=1\columnwidth]{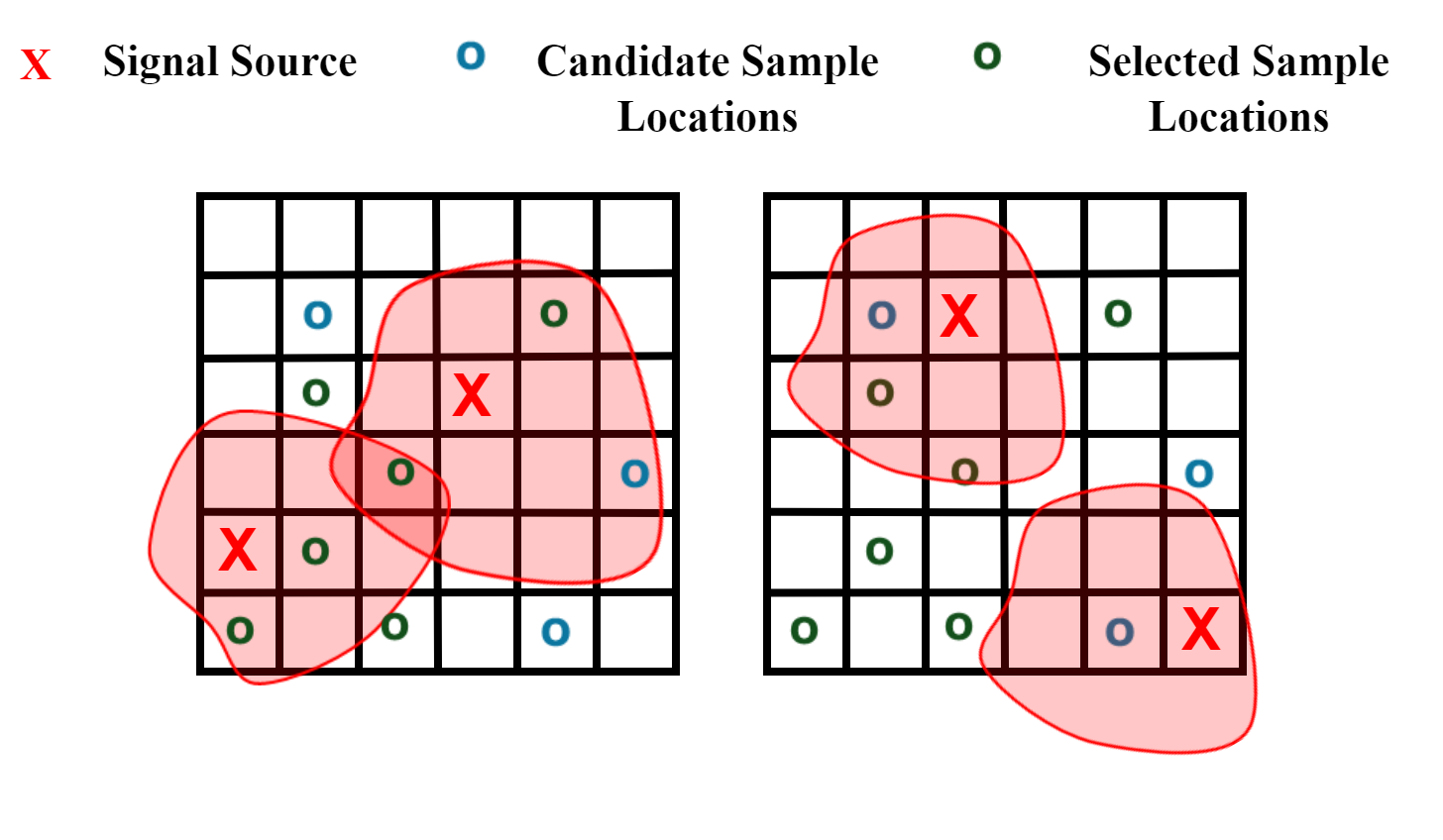}
        \caption{Visualization of signal source placement relative to the domain/geographic-region of interest. The signal source~(denoted as X) influences the candidate sample locations~(denoted as O) differently depending on its position, altering the source coverage boundary (denoted in green color).}
        \label{fig: Methodology}
%        \vspace{-13pt}
\end{figure}

Here, we focus on developing the spatial REM of a specific geographical region of interest denoted as Domain~\(\mathbb{D}\). The Domain~\(\mathbb{D}\) is discretized into $I \times J$ equally sized cells, with each cell %pixel
representing a spatial location within the region of interest. 
We assume the presence of a transmitter~$X$ which may be located within or outside~\(\mathbb{D}\), as illustrated in Fig.~\ref{fig: Methodology}. 

\begin{table}[!b]
\centering
\caption{Notation Summary}
\label{tab:notation}
\small % Reduce font size for the table
\begin{tabular}{|p{1.5cm}| p{6cm}|}\hline
\renewcommand{\arraystretch}{1.4}
\textbf{Notation} & \textbf{Description} \\ \hline\hline
$\mathbb{D}$ & Domain representing the specific geographical region of interest \\ \hline
$I \times J $ %$$\mathbb{I} \times \mathbb{J}$ 
& Grid dimensions of the domain $\mathbb{D}$ \\ \hline
%$\mathbb{T}$ & Transmitter operating within or Outside of the domain $\mathbb{D}$ \\ \hline
$\Omega_n$ & Set of RF sensors sparsely deployed across the region $\mathbb{D}$; $n = 1,2, \ldots, N$ \\ \hline
$P^{\text{obs}}(\Omega_n, C)$ & Observed power measurements (RSSI) at sensor locations $\Omega_n$ for channel $C$ \\ \hline
$\mathbb{C}$ & Set of channels or spectrum of interest, partitioned into discrete frequency bands \\ \hline
$P^{\text{pred}}$ & Predicted power measurements (RSSI) by the model for sensor locations \\ \hline
$L$ & Objective function representing the error between observed and predicted values \\ \hline
\end{tabular}
\end{table}
The transmitter's location and other parameters (e.g., transmission power and antenna characteristics) %the underlying PSD 
are unknown. % to the modeling engine. 
\iffalse 
\begin{equation}
    \Omega_n = \{ (i, j) \mid i \in [I], \; j \in [J] \} \subseteq \mathbb{I} \times \mathbb{J}
    \label{eq:sensor_set}
\end{equation}
\fi 
A set of RF sensors \mbox{$\Omega_n$ ($n = 1, 2,  \ldots, N$)} %, denoted in Equation~\eqref{eq:sensor_set}, 
is sparsely deployed across $\mathbb{D}$ at random to observe the Received Signal Strength Indicator~(RSSI) at selected locations. %in dBm. 
Each sensor provides measurements of RSSI or power levels $P^{\text{obs}}$~$(\Omega_n, C)$ for each channel~$C$ in a given set~$\mathbb{C}$. The set of channels~$\mathbb{C}$ represents the spectrum of interest, partitioned into discrete frequency bands, with each band corresponding to a unique channel. The measurements collected by these sensors are sparse and are impacted by shadowing effects and large-scale path loss, which vary across both spatial and spectral dimensions. 

Our primary objective in SC %radio environment mapping 
is to model a function that can accurately predict the RSSI at any  location within~$\mathbb{D}$, enabling the generation of the REM for the domain~$\mathbb{D}$.  Mathematically, the objective is to minimize the error~($L$) between the expected observed RSSI~($P^{\text{obs}}$) and the expected model-predicted RSSI~($P^{\text{pred}}$). The notations used in this paper are summarized in TABLE~\ref{tab:notation}.
\begin{equation}
    L = \frac{1}{N} \sum_{n=1}^{N} | {P}^{\text{pred}}(\Omega_n, C) -P^{\text{obs}}(\Omega_n, C) |. 
    \label{eq:objective}
\end{equation}

\subsection{Physics-Informed Neural Network in ReVeal-MT}

Physics-informed neural networks are a recent development in scientific machine learning that leverage the ability of neural networks to learn underlying physics. The core idea behind PINNs is to incorporate the governing physical equation, typically a PDE, as a component of the neural network's loss function during training. The mean squared residual of the governing PDE, along with the data-driven loss function, is minimized to train the %fully connected 
neural network effectively. 

\iffalse 
PINN approaches to approximate the solution of the problem through a representation model defined as
\begin{equation}
    \hat{u}(\Omega_n) \approx u(\theta, \Omega_n), \Omega_n \in \mathbb{D}.
    \label{eq:model}
\end{equation}
\noindent Here, $u(\theta, \Omega_n)$ denotes the representation model with $\theta$ being the learnable parameters, 
% where $x$ denotes the spatial coordinates, 
and $\hat{u}(\Omega_n)$ represents the solution to the ODE/PDE. 
\fi

%\subsection{Network Architecture}

\begin{figure}[!htbp]
        \centering
        \includegraphics[width=1\columnwidth]{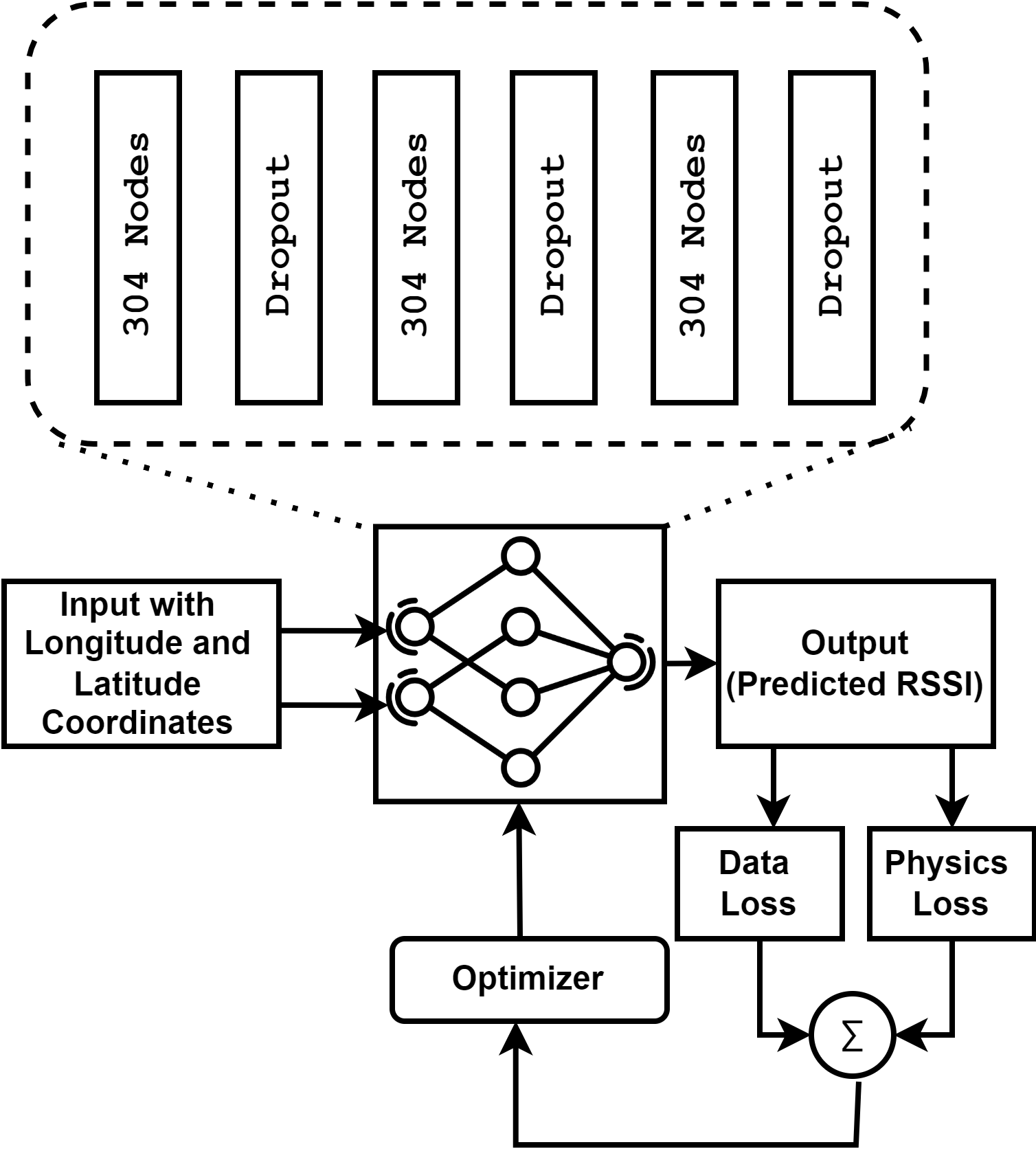}
        \caption{PINN architecture of ReVeal}
        \label{fig: Pinn_architecture}
%        \vspace{-13pt}
\end{figure}

To solve the SC problem, % defined in Equation~\eqref{eq:objective}, 
we propose a PINN architecture as illustrated in Fig.~\ref{fig: Pinn_architecture}.
The network input consists of a set of spatial locations of the measurement points $\Omega_n$ representing the geographical locations %(i.e., the latitude and longitude coordinates) 
of the RF sensors within the Domain~$\mathbb{D}$. For a given channel~$C$, the output of the PINN $P^{\text{pred}}$ represents the expected RSSI values at the specified locations.

During training, $P^{\text{pred}}$ is compared with the observed values at each training sample to compute the data-driven loss. Apart from the data-driven loss, the residual of the governing-physics PDE is incorporated to further minimize the overall error of the PINN. The calculated loss is fed into the optimizer and is used to update %$\theta$, 
the network's weights and biases accordingly. % layers. 
To prevent overfitting, multiple dropout layers are included in the neural network architecture.

The overall loss function of the PINN ensures that the model not only minimizes the Mean Absolute Error~(MAE) based on the measured data, but also adheres to the underlying physics, helping the model to generalize better during the training process. The overall loss function for the PINN optimization can be defined as

\begin{equation}
L_{\text{total}} = (1- \lambda) L_d + \lambda L_p,
\label{eq:total}
\end{equation}
where
\begin{comment}
\begin{equation}
L_d = \frac{1}{N} \sum_{i=1}^{N} \left| f(x_i) - y_i \right|^2,
\label{eq:Ld}
\end{equation}
and
\begin{equation}
L_p = \frac{1}{N} \sum_{i=1}^{N} \left| \nabla^2 \hat{u}(\mathbf{x}_i) \right|^2.
\label{eq:Lp}
\end{equation}
\end{comment}
\begin{equation}
L_d = \frac{1}{N}  \sum_{n=1}^{N} \left| {P^{\text{pred}}}(\Omega_n, C) -{P^{\text{obs}}}(\Omega_n, C) \right|
\label{eq:Ld}
\end{equation}
and 

\begin{align}
R(\Omega_n, C) &= \nabla^2 P^{\mathrm{pred}}(x,y) \nonumber\\
& - \Bigg[\,10 \sum_{i=1}^M w_i(x,y)\,\nabla^2 a_i(x,y) \nonumber\\
& + 10\ln 10 \Bigg(\sum_{i=1}^M w_i(x,y)\|\nabla a_i(x,y)\|^2 \nonumber \\
& - \Big\|\sum_{i=1}^M w_i(x,y)\nabla a_i(x,y)\Big\|^2\Bigg)\,\Bigg].
\label{eq:residual_mt}
\end{align}

\begin{equation}
L_p = \frac{1}{N}\sum_{n=1}^{N} \big|\, R(\Omega_n, C)\,\big|.
\label{eq:Lp}
\end{equation}

\noindent Here, $L_d$ refers to the loss calculated from the labeled data points collected from the sensors. The physics-driven loss $L_p$ is defined by Eqn.~\eqref{eq:Lp}, where $\nabla^2$ represents the Laplacian of both the predicted RSSI and observed RSSI at the sample location (a detailed analogy of introducing $L_p$ is discussed in the next sub-section. The parameter~$\lambda$ in Eqn.~\eqref{eq:total} is a variable that controls the weightage given to the data-driven or physics-driven loss during the training of the neural network.

\subsection{Governing-Physics PDE in ReVeal-MT}

A key design decision in ReVeal-MT is selecting the physics model that governs the spatial dynamics of RSSI when multiple transmitters are transmitting in the Domain $\mathbb{D}$. To achieve this, we derive a second-order PDE based on well-established wireless signal path loss models. 
    Specifically, we extend the well-known single-transmitter path-loss model to a multi-transmitter path-loss model and the partial derivative of the governing equations. In a multi-transmitter scenario, the pathloss model is given by
\begin{align}
P_{r,i}^{\mathrm{dB}}(x,y) = P_{T,i}^{\mathrm{dB}} & - 10 \eta \log_{10}\Big(\frac{r_i(x,y)}{d_0}\Big)\nonumber\\ &+ Z_{\sigma,i}(x,y)
\label{eqn:path_loss}
\end{align}
Eqn.~\eqref{eqn:path_loss} describes the received signal power $P_{r,i}^{\mathrm{dB}}(x,y)$ at any given instance in space $(x, y)$ from transmitter $i$ in \textbf{dB scale}, as a function of the transmit power $P_{T,i}^{\mathrm{dB}}$, distance from the transmitter $r_i(x,y) = \sqrt{(x-x_i)^2 + (y-y_i)^2}$, path loss exponent $\eta$, reference distance $d_0$, and an added shadowing factor $Z_\sigma$ denoting a specific realization of the underlying random spatial variations\footnote{Temporal variation caused by fading is beyond the scope of this study and is left as a part of the future work.}. %Since the received signal is affected due to the added randomness by \( Z_\sigma \), 

Now, in the case of a multi-transmitter scenario, the received signal strength from multiple transmitters in the dB-scale cannot be simply added, so for summing contributions from multiple transmitters, we convert dB to linear scale using
\begin{equation}
P_{r,i}^{\mathrm{linear}}(x,y) = 10^{P_{r,i}^{\mathrm{dB}}(x,y)/10}.
\label{eqn:dB_to_linear}
\end{equation}
Substituting the expression for $P_{r,i}^{\mathrm{dB}}(x,y)$ in Eqn.~\eqref{eqn:dB_to_linear} gives:

\begin{equation}
P_{r,i}^{\mathrm{linear}}(x,y) = 10^{\frac{1}{10} \Big( P_{T,i}^{\mathrm{dB}} - 10 \eta \log_{10} \frac{r_i(x,y)}{d_0} + Z_{\sigma,i}(x,y) \Big)}.
\label{eqn:linear}
\end{equation}
For compact notation, we define:
\begin{equation}
a_i(x,y) = \frac{1}{10} \Big( P_{T,i}^{\mathrm{dB}} - 10 \eta \log_{10} \frac{r_i(x,y)}{d_0} + Z_{\sigma,i}(x,y) \Big),
\end{equation}
so that the linear received power simplifies to
\begin{equation}
P_{r,i}^{\mathrm{linear}}(x,y) = 10^{a_i(x,y)}.
\end{equation}
The total received power at a specific receiver location in linear scale from all transmitters is defined by: 
\begin{equation}
S(x,y) = \sum_i P_{r,i}^{\mathrm{linear}}(x,y) = \sum_i 10^{a_i(x,y)}, 
\end{equation}
where $i$ represents the number of transmitters. Converting the linear expression back to $\mathrm{dB}$:
\begin{equation}
P_{r,\mathrm{tot}}^{\mathrm{dB}}(x,y) = 10 \log_{10} S(x,y) = 10 \log_{10} \Big( \sum_i 10^{a_i(x,y)} \Big).
\label{eqn:total_dB}
\end{equation}
By taking the Laplacian of the Eqn.~\eqref{eqn:total_dB} the expression reduces to 
\begin{align}
\nabla^2 P_{r,\mathrm{tot}}^{\mathrm{dB}}(x,y)
= & 10 \sum_i w_i(x,y)\,\nabla^2 a_i(x,y) \nonumber\\
& + 10\ln 10 \Bigg(\sum_i w_i(x,y)\,\|\nabla a_i(x,y)\|^2 \nonumber \\
& - \Big\|\sum_i w_i(x,y)\,\nabla a_i(x,y)\Big\|^2\Bigg), 
\label{eqn:final}
\end{align}
where the term $w_i$ is defined as 
\begin{equation}
w_i=  \frac{10^{a_i(x,y)}}{\sum_i 10^{a_i(x,y)}}.
\end{equation}
Eqn.~\eqref{eqn:final} is the generic form for a multi-transmitter use-case. A single transmitter is a special case of this equation, and when $i=1$, the Eqn.~\eqref{eqn:final} reduces back to\footnote{See Appendix for the detailed multi-transmitter derivation.}: 
\begin{equation}
\nabla^2 P_{r,\mathrm{tot}}^{\mathrm{dB}} = 10 \sum_i w_i \nabla^2 a_i,
\label{eqn:single_transmit}
\end{equation}
which as derived in the original work~\cite{ReVeal} will reduce to
\begin{equation}
\frac{\partial^2 [P_{r,\mathrm{tot}}^{\mathrm{dB}}]}{\partial x^2} + \frac{\partial^2 [P_{r,\mathrm{tot}}^{\mathrm{dB}}]}{\partial y^2} = \frac{\partial^2 [Z_\sigma]}{\partial x^2}+\frac{\partial^2 [Z_\sigma]}{\partial y^2}.
\label{eq:sum}
\end{equation}

\noindent The left hand side of Eqn.~\eqref{eq:sum} represents the Laplacian operator as defined in Eqn.~\eqref{eq:Lp} while the right-hand side captures variations in signal strength due to shadowing caused by heterogeneous terrain and buildings within in the domain. The PDE constraint in Eqn.~\eqref{eq:sum} explicitly encodes the spatial behavior of shadowing without requiring the knowledge of the transmitter's location, ensuring that the single-transmitter use-case of ReVeal does not merely fit the training data but also generalizes effectively in unseen regions.

The PDE constraint in Eqn.~\eqref{eqn:final} encodes the spatial behavior of the received signal strength when multiple transmitters are operating in overlapping coverage regions.
This presents a fundamental challenge not present in the single-transmitter case described by Eq.~\eqref{eq:sum}. In that single-transmitter scenario, the effects of transmission and shadowing on signal variation can be learned without prior knowledge of transmitter location or power. In contrast, the multi-transmitter PDE reveals that the Laplacian of the signal strength is a direct function of these critical parameters: the relative distance from each transmitter and its individual transmit power. This raises the challenge to treat them as trainable parameters and learn about those parameters during the training process.

In homogeneous environmental conditions or simple free-space settings, the signal variation due to shadowing is zero, thus the contributions of shadowing in Eqn.~\eqref{eqn:final} remains zero and does not vary across space, as a result the aggregate received signal strength at any given point in domain $D$ will be dependent on the individual contribution from each transmitter.   
However, in real-world environments, where shadowing effects are present, the right-hand side of Eqn.~\eqref{eqn:final} becomes non-zero. Thus, a good model must precisely capture the impact of shadowing while finding the  contributions from each transmitter. The physics informed loss term $L_p$ as defined in Eqn.~\eqref{eq:Lp} ensures that model not only adheres to the data collected from the field but also adheres to the physical phenomenon defined in Eqn.~\eqref{eqn:final}.

\subsection{Learning Criteria and Algorithm of ReVeal-MT}

During training, the network optimizes a  composite loss function that integrates both data-driven and physics-driven losses. After the forward pass, the model predicts the expected RSSI at the training sample %collocation 
points. The model then computes the error between the predicted and observed values using the data-driven loss defined in Eqn.~\eqref{eq:Ld}. In addition, leveraging the network's automatic differentiation capability, the model calculates the second-order spatial derivatives of the predicted RSSI values, which are used to evaluate the residuals of the governing-physics PDE defined in Eqn.~\eqref{eq:Lp}. The physics-driven loss term minimizes these residuals, ensuring consistency with the governing-physics PDE relative to the observed RSSI. % equation described in Equation~\eqref{eq:sum}. 
During back-propagation, the optimizer adjusts the network’s weights and biases to minimize the composite loss, effectively balancing both data-driven and the physics-driven loss components. 

The trade-off between the physics-driven loss and the data-driven loss is controlled by the parameter~$\lambda$, which determines their relative importance. 
Without the data-driven loss term, the model lacks a starting point for optimization, as the physics-driven loss term, based on a PDE, does not provide sufficient guidance to align predictions with real-world observations. Therefore, an appropriate choice of~$\lambda$ is essential to achieve an optimal balance between empirical accuracy and physical consistency during training, as further  elaborated  in Section~\ref{sec:results}.

\subsection{Optimizing the PINN Architecture}

Hyper-parameter tuning is a crucial yet tedious task in designing machine learning algorithms, including PINNs. The optimal selection of parameters---such as the number of hidden layers, number of neurons per layer, activation function, and learning rate---significantly impacts both the performance and convergence of the model. In literature, advanced hyper-parameter tuning algorithms utilizing techniques such as grid search and random search have been employed for parameter selection, including Autotune~\cite{Autotune} and SMAC~\cite{SMAC}. However, most of such libraries require a predefined search space from the user to identify the best parameters that minimize an objective function. On the other hand, libraries such as Optuna~\cite{optuna} provide users with the flexibility to define a dynamic search space and employ advanced optimization techniques such as Tree-Structured Parzen Estimator (TPE) for dynamic and efficient hyper-parameter tuning and optimal parameter selection. Therefore, in this study, we use Optuna~\cite{optuna} as the hyper-parameter optimization library to select parameters based on the spatially sampled data points. The resulting optimized hyper-parameters are presented in TABLE~\ref{tab:hyperparameters}. Bringing together all the aforementioned design choices, Algorithm~\ref{alg:ReVeal-MT} summarizes the ReVeal-MT algorithm.

% Learning rate                    & 0.003691329633424969         \\ \hline

\begin{table}[ht]
\centering
\caption{Hyper-parameters of ReVeal-MT}
\label{tab:hyperparameters}
\begin{tabular}{|l|l|}
\hline
\textbf{Hyper-parameter}         & \textbf{Value}               \\ \hline\hline
Number of input features         & 2                            \\ \hline
Number of hidden layers          & 3                            \\ \hline
Number of neurons per layer      & 304                          \\ \hline
Activation function              & ReLU                         \\ \hline
Dropout rate                     & 0.2                          \\ \hline
Number of output features        & 1                            \\ \hline

Learning rate                    & 0.00369                               \\ \hline

\end{tabular}
\end{table}

\begin{algorithm}[!t]
\caption{\textbf{ReVeal-MT for Multi-Transmitter REM Generation}}
\label{alg:ReVeal-MT}
    \KwData{Input: Sensor data $\{\Omega_n, P^{\text{obs}}(\Omega_n)\}$, domain $\mathbb{D}$, number of transmitters $M$, path-loss exponent $\eta$, reference distance $d_0$}
    \KwResult{Radio Environment Map (REM)}
    \textbf{[Initialize Model and Parameters]}\; 
    $model = \text{PINN}(\text{hyper-parameters})$ \;
     \textbf{[Initialize trainable transmitter parameters]}\;
    Powers: $\{P_{T, i}\}_{i=1}^M$\;
    Locations: $\{(x_i, y_i)\}_{i=1}^M$\;
    
    \textbf{[Define the Loss Functions]}\;
    $L_d = \text{MAE}(P^{\text{pred}}(\Omega_n), P^{\text{obs}}(\Omega_n))$ \tcp{Data loss}
    $L_p = \frac{1}{N} \sum_{n=1}^{N} |R(\Omega_n,C)|$\tcp{Physics loss (PDE residual from Eqn. (13))}
    $L_{\text{total}} = (1-\lambda) L_d + \lambda L_p$\tcp{Weighted total loss}
    
     \textbf{[Training Loop]}\;
    \For{\textit{epoch} = 1 \textbf{to} \textit{num\_epochs}}{
        \For{\textit{batch} \textbf{in} \textit{batches}}{
            Predict $P^{\text{pred}}, Z_{\sigma} = model(\textit{batch})$\tcp{Forward pass}
            Compute $L_d$ \tcp{Compare prediction to observation}
            Compute $L_p$ \tcp{Using current transmitter parameters}
            $loss = L_{\text{total}}$\;
            Update $model$ parameters \& transmitter parameters using $optimizer.step(loss)$
        }
    }
    
    \textbf{[Generate Final Output]} \;
    $\text{REM} = model(\mathbb{D})$ \tcp{Predict over the entire domain}
    \Return $\text{REM}, \{P_{T, i}\}_{i=1}^M, \{(x_i, y_i)\}_{i=1}^M$ \tcp{Return map and transmitter estimates}
\end{algorithm}

\section{Experimental Setup} % and Evaluation}
\label{sec:experimental_setup}

\begin{figure}[!t]
        \centering
        \includegraphics[width=1\columnwidth]{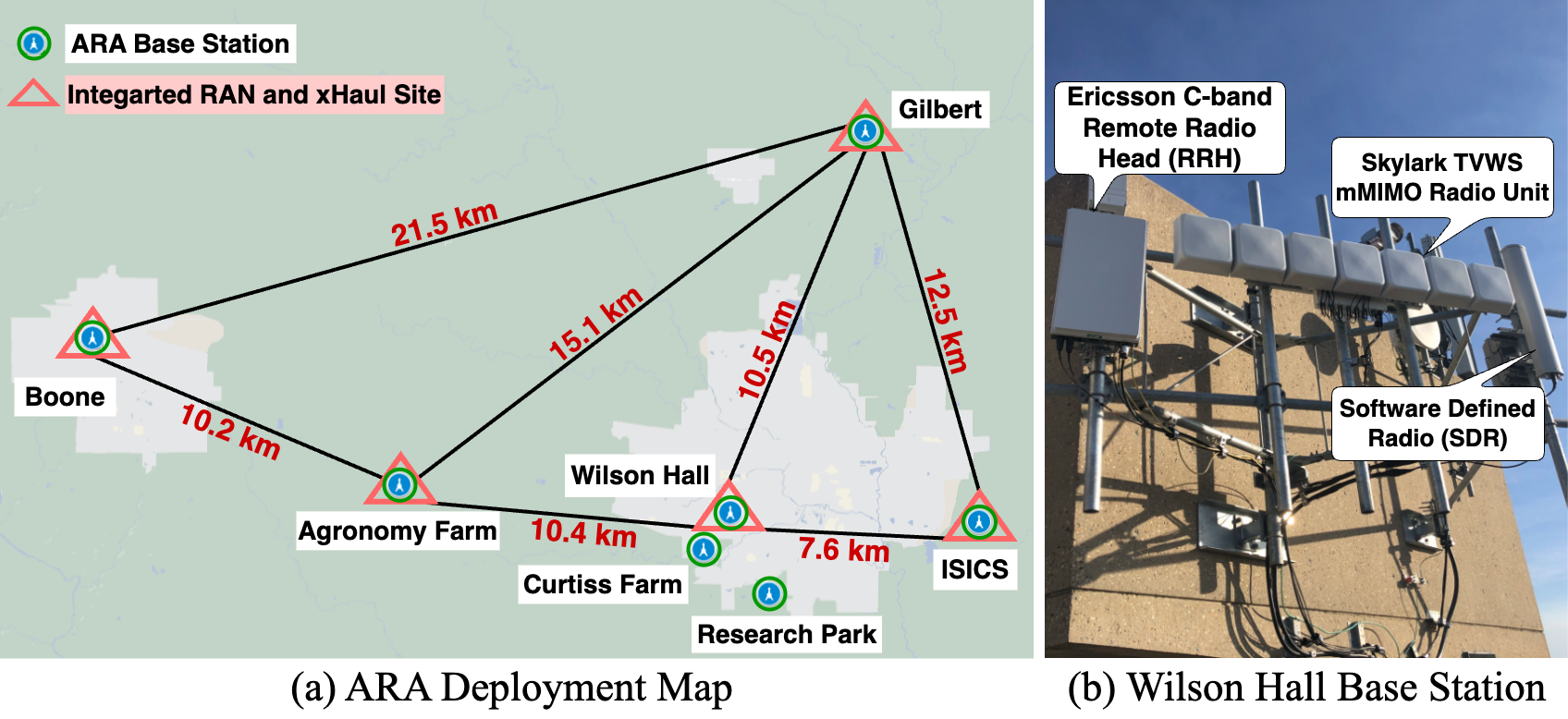}
        \caption{ARA Deployment 
        }
        \label{fig: ARA_deployment}
%        \vspace{-13pt}
\end{figure}

We evaluate ReVeal-MT using real-world data collected from the ARA testbed \cite{ARA_design_implementation}. ARA is a first-of-its-kind wireless living lab located around Iowa State University, spanning an area over 500 square kilometers and covering research and producer farms along the rural communities of Central Iowa. Fig.~\ref{fig: ARA_deployment} 
illustrates the deployment of seven ARA Base Station~(BS) sites. Among these, Wilson Hall, Boone, ISICS, and Gilbert are equipped with the SkyLark massive MIMO~(mMIMO) platform operating in the TVWS band. Meanwhile, Wilson Hall, Agronomy Farm, Research Park, and Curtis Farm sites include a C-Band Commercial-Off-the-Shelf~(COTS) mMIMO system from Ericsson operating at 3450--3550\,MHz band. In addition, all BS sites feature Software Defined Radios~(SDRs) operating at 3400--3600\,MHz, supporting fully programmable,  end-to-end, whole-stack 5G experiments using open-source systems such as OpenAirInterface and srsRAN.

\paragraph{Data Collection Site}

\begin{figure}[!htbp]
        \centering
        \includegraphics[width=.8\columnwidth]{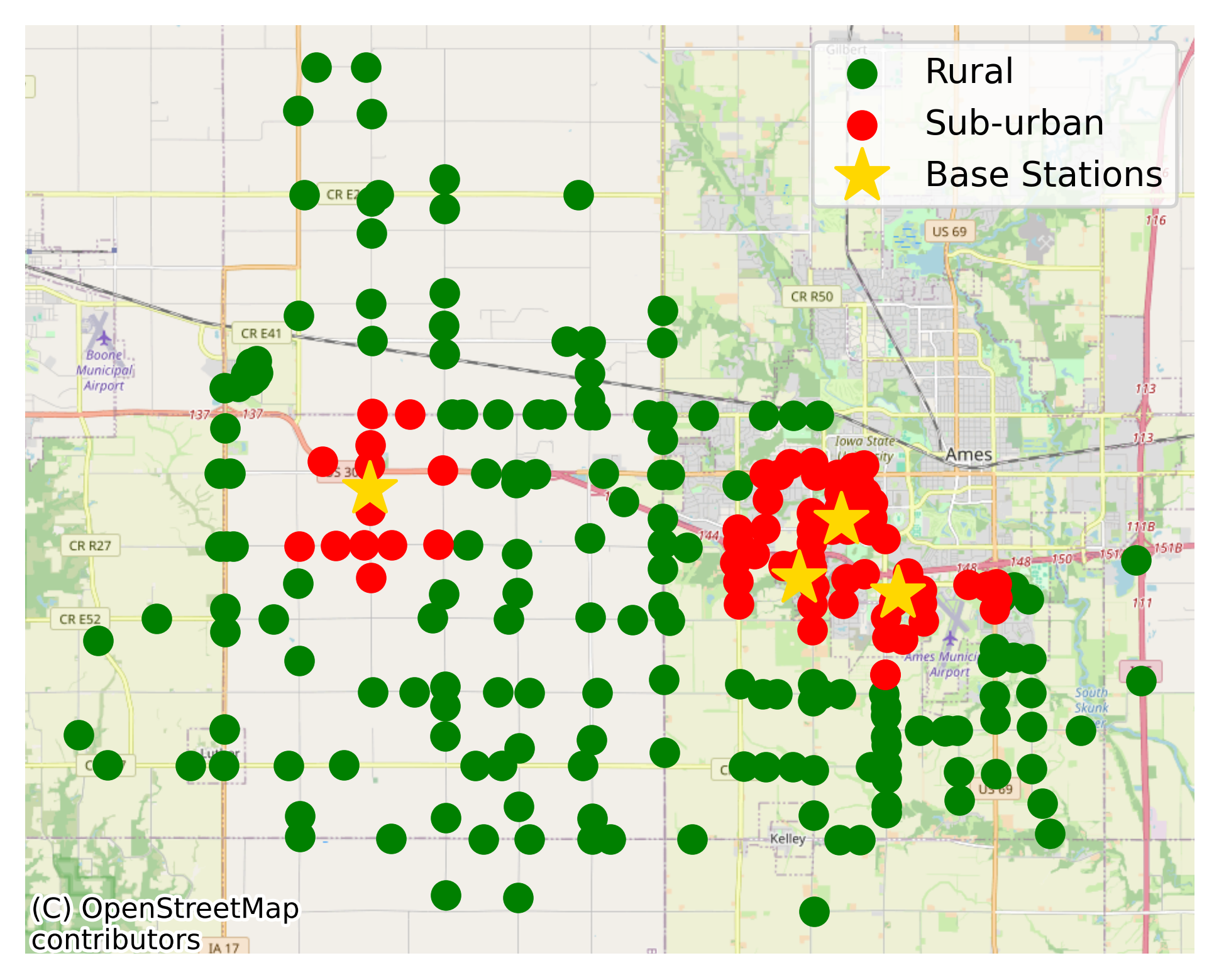}
        \caption{Sampling Locations and Corresponding Terrain Conditions }
        \label{fig:dataSamples}
%        \vspace{-13pt}
\end{figure}

\begin{figure}[!htbp]
        \centering
        \includegraphics[width=0.8\columnwidth]{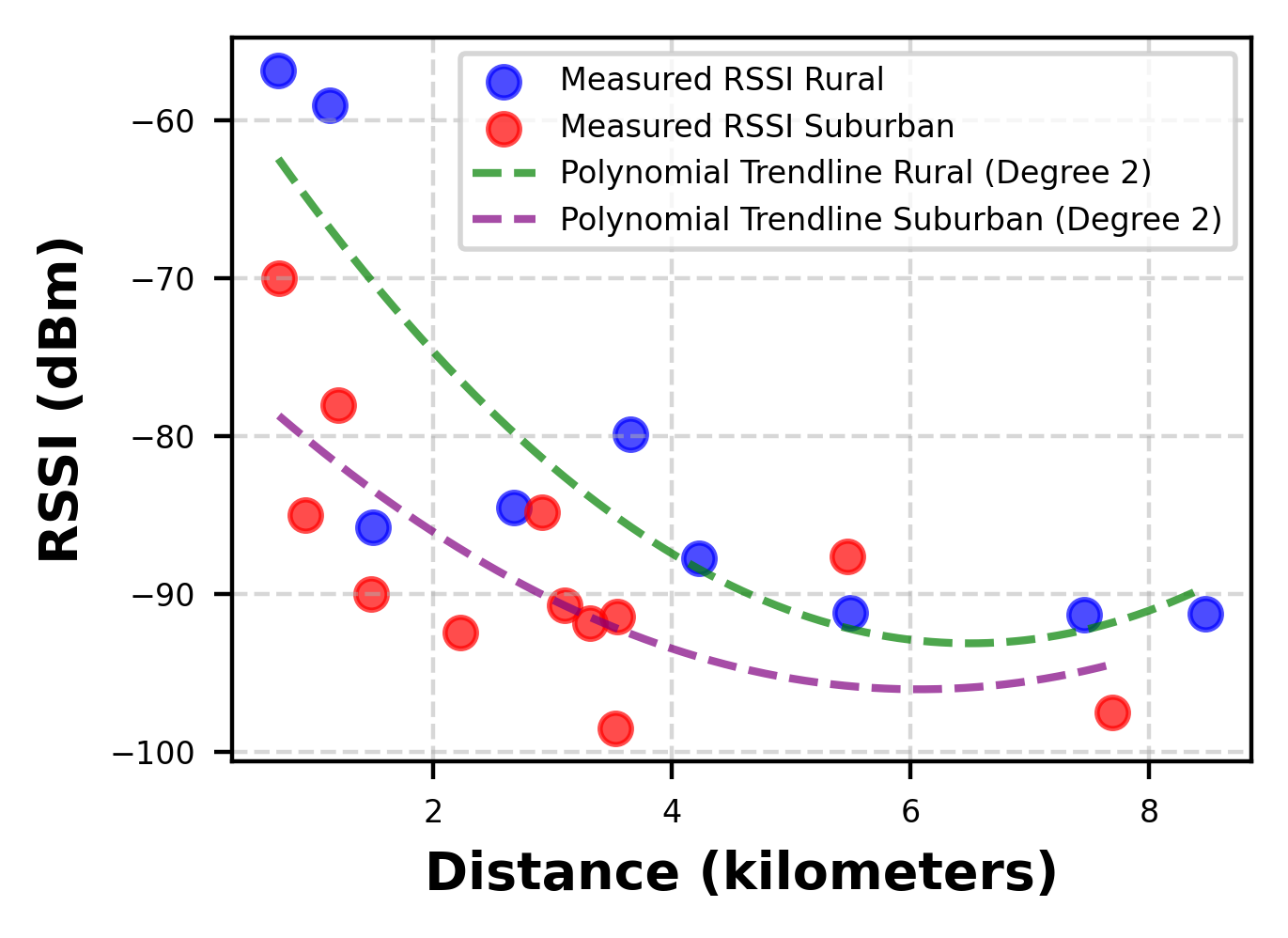}
        \caption{Variation in Channel Conditions across Rural and Suburban Terrains 
        }
        \label{fig:ruralSurburbanChannels}
%        \vspace{-13pt}
\end{figure}

For REM modeling, we collected real-world data using the Ericsson mMIMO systems deployed at the Wilson Hall, Curtis Farm, Research Park and Agronomy Farm Base-Stations. The sample points were selected around all 4 base-stations covering an area spanning 370 square kilometers through a drive test. During the drive test almost 65,000 spatial sample points were collected in the given area and out of the collected samples we used selected number of samples using the Local Pivotal Method described in the next sections.
Before initiating the actual sampling process, we ensured that no other transmitters were operating at the same frequency as the C-Band band used by the Ericsson BS. % within the pre-defined computational domain. % during the pre-sampling phase. 
The coverage range of the ARA BSes %and Ericsson BSes 
encompasses diverse %variety of 
terrains, ranging from rural communities to suburban areas of Downtown Ames, as illustrated in Fig.~\ref{fig:dataSamples}. These different terrains exhibit %are impacted by 
varying distributions of shadowing and fading effects. % due to the diverse channel conditions in diverse terrains. 
As shown in Fig.~\ref{fig:ruralSurburbanChannels}, the measurement sampling instances indicate different levels of shadowing, even when the distance from the BS remains the same.


\begin{figure}[!htbp]
        \centering
        \includegraphics[width=0.8\columnwidth]{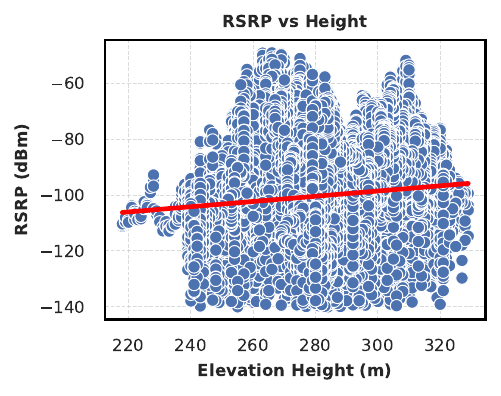}
        \caption{ Scatter plot of RSSI versus elevation, showing the correlation between received signal strength and measurement height 
        }
        \label{fig:distance_correlation}
%        \vspace{-13pt}
\end{figure}

Furthermore, Fig.~\ref{fig:distance_correlation} visualizes the relationship between received signal strength (RSSI) and elevation, highlighting that terrain-induced shadowing causes significant RSSI variation even at comparable distances from the transmitter, underscoring the need for environment-aware modeling.

\paragraph{Spectrum Sensing Equipment}

The ARA BS sites are equipped with spectrum sensing equipment, specifically the Keysight N6841-A RF sensors %from Keysight installed at phase 1 BS locations. These spectrum sensors are 
connected to Keysight N6850A omni-directional antennas, which monitor spectrum activities across the bands of interest. Apart from the fixed spectrum sensors, the Keysight N9952A FieldFox, equipped with an N6850A omni-directional antenna, and the Keysight NEMO Handy handheld measurement solution are used to capture RSSI values at various spatial points around the BS site.

\paragraph{Spatial Sampling Strategy}

Selecting appropriate spatial samples %from candidate points 
is crucial to reducing sampling overhead while maintaining modeling accuracy. A spatially-balanced sampling strategy is essential in spatial modeling to minimize prediction errors across the entire domain. The primary objective of this approach is to identify the most informative sample points from all the candidate candidate locations in the dataset~\cite{spatial_sampling}. 

Since spatial points in close proximity often exhibit similar or identical RSSI values, it is preferable to select spatially distant points that are representative of the overall population. To this end, we employ Local Pivotal Method~(LPM)~\cite{grafstra2012spatially} to determine the spatial sampling locations in our study, as shown in Fig.~\ref{fig:dataSamples}. % and the LPM serves that purpose best. 
\iffalse 
\red{Using the function \texttt{lpm(.)} from the \texttt{Balancedsampling} package in the \texttt{R} programming language, samples of varying sizes (such as 40, 60, and 80) are drawn. Then, based on the predictive performance of our method, we can decide how much we can downscale the actual number of samples needed.} 
\hz{1) It will be useful to present the sampling analysis results and show how the specific set of locations are decided. 2) Exactly what are the sampling points used in the paper?}
\fi

\begin{table*}
\centering
\caption{Performance of Different Path Loss Models Across the Whole Domain
}
\label{tab: metrics}
\normalsize % Reduce font size for the table
\begin{tabular}{|p{6cm}|c|c|c|c|}\hline
\renewcommand{\arraystretch}{1.4}
\textbf{Model} & \textbf{RMSE (dB)} & \textbf{MAE (dB)} & \textbf{R-Squared} & \textbf{Computation Time (seconds)} \\ \hline\hline
3GPP TR 38.901 Model & 20.16 & 19.85 & -0.68 & 3.3 \\ \hline
ITU-R IMT-2020 Model & 14.13 & 12.35 & -0.16 & 3.5 \\ \hline
Ray-Tracing with Sionna & 32.93 & 26.04 & 0.30 & - \\ \hline
Kriging & 18.02 & 16.42 & -0.42 & 39 \\ \hline
FCNN & 11.95 & 9.86 & 0.24 & 42.6 \\ \hline
ReVeal & 9.75 & 7.80 & 0.35 & 35.6 (With early stopping)\\ \hline
ReVeal-MT & 2.66 & 2.20 & 0.84 & 45.6 (With early stopping)\\ \hline
\end{tabular}
\end{table*}

\section{Experimental Results}
\label{sec:results}

The computational experiments for evaluating ReVeal-MT were conducted on a workstation equipped with an Intel® Xeon® processor operating at 3.40\,GHz with 32\,GB DDR4 RAM. ReVeal-MT is compared against %different benchmarks available from 
stochastic models, including %such as
3GPP TR 38.901~\cite{3gpp_ts_38.901} and ITU-R IMT-2020~\cite{etsi_tr_138901}. Deterministic models such as ray tracing using Sionna~\cite{sionna} is also implemented for the specific ARA regions where real-world measurement data is collected. 
Furthermore, ReVeal-MT is benchmarked against classical machine learning models commonly used for generating REMs, including the kriging and standard FCNNs. 
To assess the importance of using the PDE-based path loss model in ReVeal-MT, we also compare it to that incorporate alternate physics models,  specifically 3GPP TR 38.901 and ITU-R IMT-2020. These variants are denoted as PINN with 3GPP TR 38.901 Model and PINN with ITU-R IMT-2020 Model, respectively.

TABLE~\ref{tab: metrics} presents the performance of various benchmark algorithms used for channel modeling. The table includes several key metrics such as Root Mean Squared Error~(RMSE), Mean Absolute Rrror~(MAE), R-squared, and computation time for each model. 
ReVeal-MT achieves an RMSE of the order of magnitude smaller than other methods while maintaining a relatively low computation time. 
    Although the statistical models of 3GPP TR 38.901 and ITU-R IMT-2020 have low computation times of 3.3\,s and 3.5\,s, respectively, they exhibit high modeling errors, % 
    with RMSE values of 20.16\,dB and 14.13\,dB, respectively. The low computation complexity but high prediction errors in these statistical models result from their reliance on predefined formulas that capture statistical signal propagation behavior without incorporating site-specific environmental features (e.g., vegetation, trees, and buildings) present in the ARA testbed during evaluation.

The computation time for data-driven models such as Kriging and fully connected neural networks (FCNNs) is higher than that of traditional statistical models, but significantly lower than that of deterministic ray tracing approaches. Among all these models, ray tracing is the most computationally intensive, as it involves detailed environmental modeling and precise simulation of individual propagation paths between the transmitter and receiver. In practice, implementations such as MathWorks' ray tracing routine can take several hours to load and execute on a local machine, depending on the complexity of the outdoor scene and the number of transmitters involved \cite{MathWorks_RayTracing}. By contrast, Sionna’s ray tracing models \cite{sionna} leverage NVIDIA GPUs for acceleration, resulting in substantially faster performance. However, a major limitation of ray tracing models is their dependence on highly accurate and detailed environmental data, which is often impractical to obtain in dynamic real-world environments. Consequently, despite their physical accuracy, these models can produce higher errors when the environmental input is incomplete or outdated.

ReVeal-MT however, achieves the best trade-off, with an RMSE of 2.66 dB and MAE of 2.20 dB, coupled with a positive R² score of 0.84. Its computational complexity is acceptable for large-scale deployments with the demonstrated coverage area of more than $357 \,\text{km}^2$, especially considering its data efficiency. Notably, when compared to our prior single-transmitter PINN \cite{ReVeal} which yields an RMSE of 7.80 dB when applied to multi-transmitter data, ReVeal-MT delivers more than a threefold reduction in error. This substantial improvement underscores the robustness of the proposed multi-transmitter formulation and establishes ReVeal-MT as the most reliable and scalable approach across diverse evaluation scenarios.

\begin{figure}[!htbp]
        \centering
        \includegraphics[width=0.9\columnwidth]{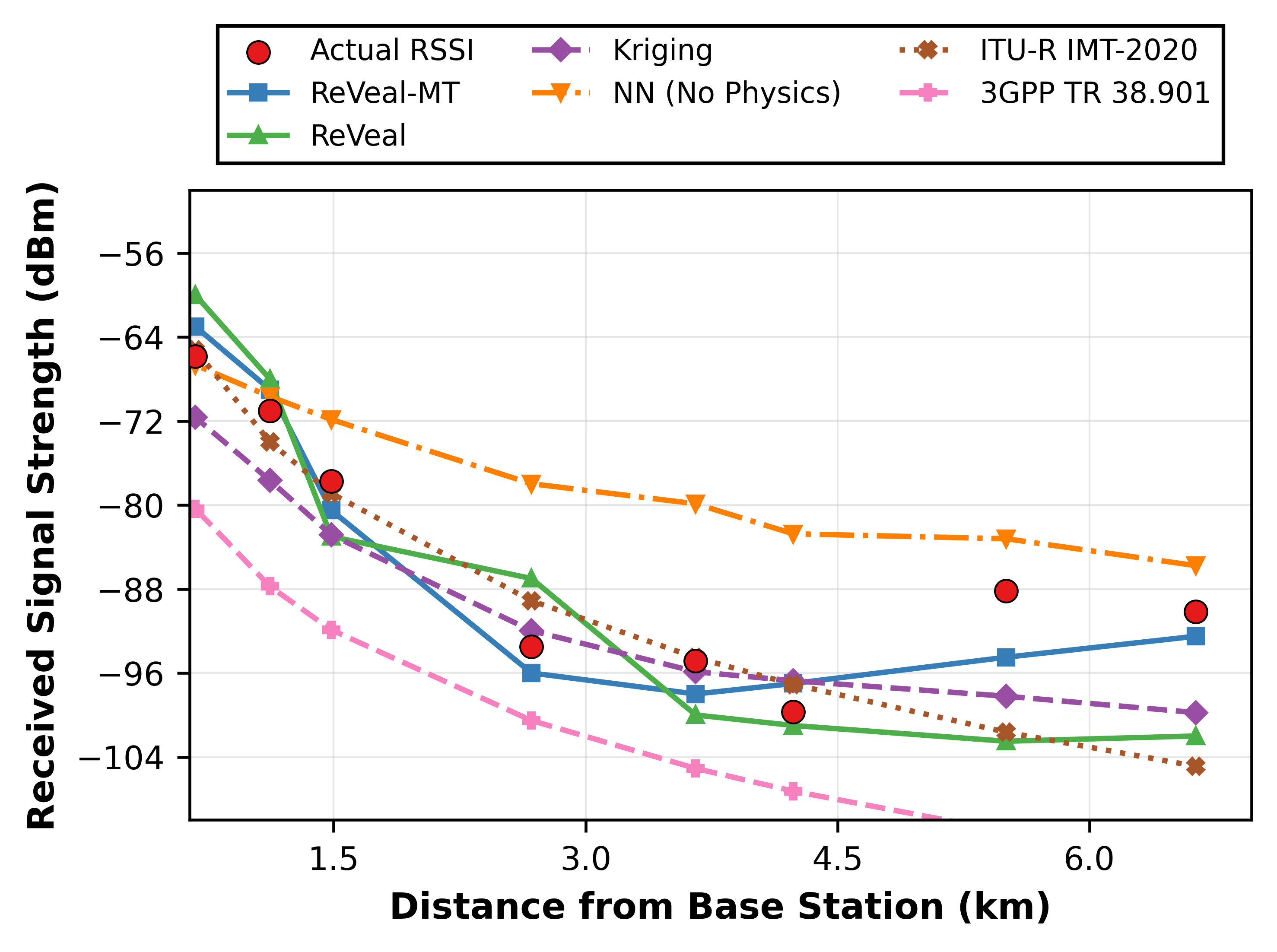}
        \caption{Comparison of Predicted and Actual RSSI Values at Different Distances from the BS 
        }
        \label{fig: RSSI_comparison}
%        \vspace{-13pt}
\end{figure}

For locations at varying distances from the Ericsson BSs, Fig.~\ref{fig: RSSI_comparison} 
compares the actual RSSI values and those predicted by different methods. 
    ReVeal-MT demonstrates strong performance across nearly all distances, effectively capturing RSSI variations caused by shadowing. In particular, ReVeal-MT closely follows the changes in RSSI influenced by  terrain conditions and over-lapping base-station coverage, showcasing its ability to model real-world signal behavior more accurately than existing methods. 
Traditional statistical %empirical 
    models tend to underestimate signal strength by tens of dBs. The kriging model exhibits higher errors in certain scenarios due to its method of estimating values at unmeasured locations by calculating a weighted average of surrounding spatial points. %, based on their spatial correlation. 
    Such an interpolation approach assumes that the underlying spatial field follows a specific correlation structure, and any deviation---such as in complex environments with significant signal strength variations---can lead to larger prediction errors. 
    In addition, PINNs with the 3GPP TR 38.901 Model and ITU-R IMT-2020 Model attempt to align with the behavior of the underlying statistical models, and % as greater weight is assigned to the physics loss. However, 
    the inherent limitations of the underlying models restrict the accuracy of the PINNs in such cases.

\begin{figure}[!htbp]
        \centering
        \includegraphics[width=1\columnwidth]{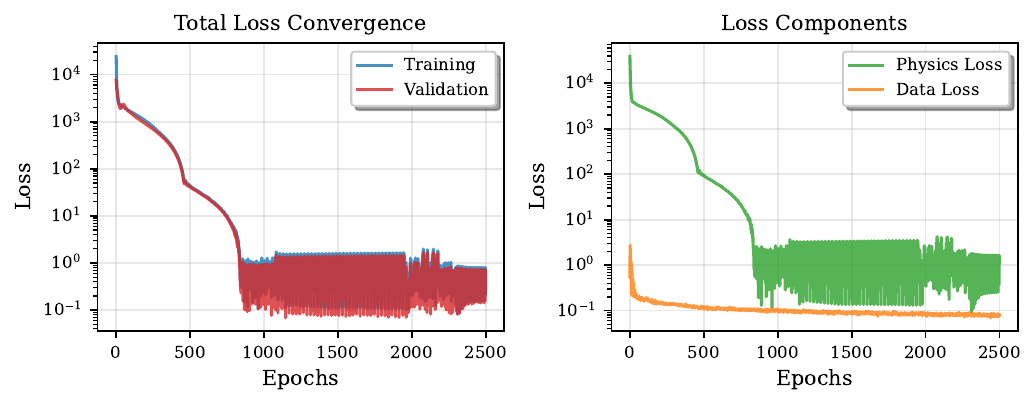}
        \caption{Comparison of Training and Test Loss for ReVeal-MT and FCNN Model 
        }
        \label{fig: Loss_coparison}
%        \vspace{-13pt}
\end{figure}

The training dynamics of ReVeal-MT are illustrated in the loss convergence plots shown in Fig.~\ref{fig: Loss_coparison}. In  Fig.~\ref{fig: Loss_coparison}~(a), both the training and validation losses decrease rapidly during the initial epochs and stabilize within approximately 1000 epochs, demonstrating fast and stable convergence. The close alignment between training and validation losses further indicates that the model generalizes well to unseen data without overfitting, a direct benefit of integrating physics-based regularization into the training process.

Fig.~\ref{fig: Loss_coparison}~(b) further decomposes the total loss into its two primary components: the data-driven loss and the physics-driven loss. The data loss steadily decreases as the model aligns with measured sensor observations, while the physics loss enforces consistency with the governing PDE across the spatial domain. The simultaneous reduction of both components highlights the complementary nature of the hybrid loss formulation, where the physics-based term prevents overfitting to sparse measurements and the data term anchors the model to empirical ground truth. Together, these curves confirm that ReVeal-MT not only converges rapidly but also achieves a balanced trade-off between empirical accuracy and physical consistency.

\begin{figure}[!htbp]
        \centering
        \includegraphics[width=1\columnwidth]{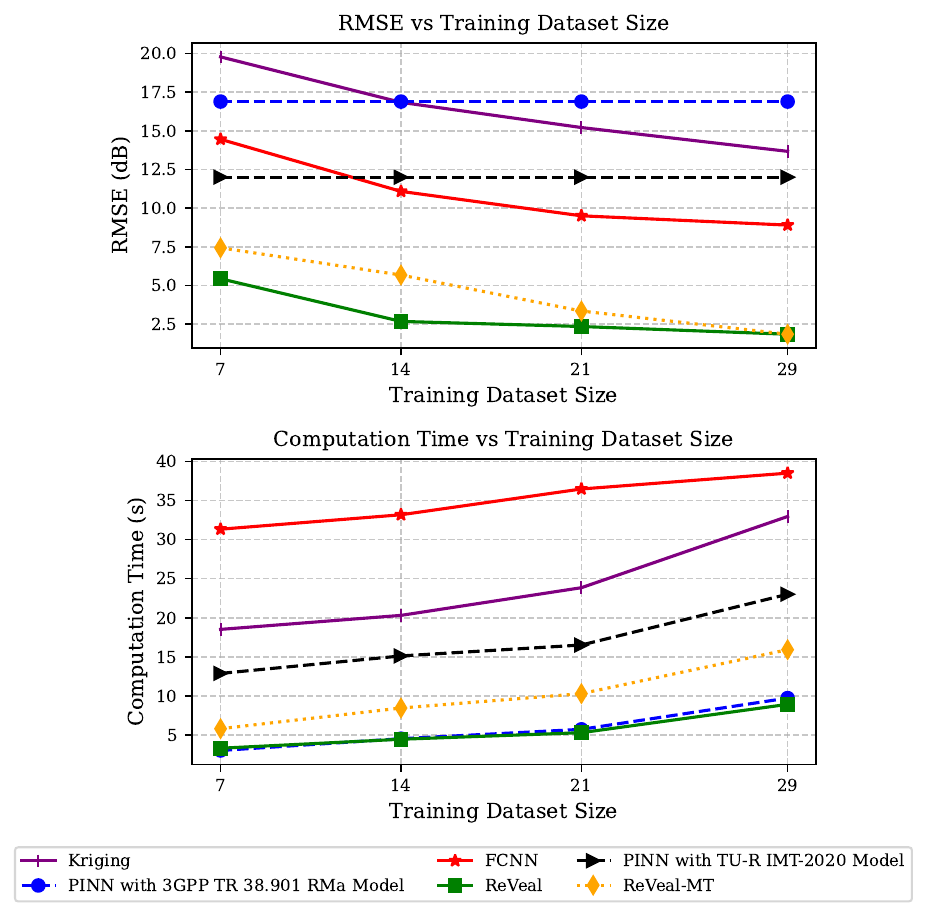}
        \caption{Performance Analysis with other Spectrum Cartography Techniques
        }
        \label{fig: Performance_Analysis}
%        \vspace{-13pt}
\end{figure}

To evaluate ReVeal-MT's capability of generating accurate REMs with sparse training samples, Fig.~\ref{fig: Performance_Analysis} 
\begin{comment}

\end{comment}
presents the performance of different data-driven models with varying sample sizes, where  PINN models (including ReVeal-MT) use $\lambda = 0.459$, as learnt by the model during the training process. 

We observe that the RMSE decreases as the number of training samples increases. % for training purposes. 
However, PINN models that utilize statistical models as their underlying physics framework hardly show any improvement, as they tend to replicate the behavior of the statistical model itself. With 14 spatial samples, the average RMSE for ReVeal-MT was slightly over 6\,dB, whereas it decreased to 2.21\,dB with  29~training samples.
As expected,  computation complexity increases with the number of training samples. 
By employing early stopping, where training halts once the desired accuracy is reached, ReVeal-MT could be trained and  used to visualize REM in approximately 45.6~seconds. In contrast, other methods require for instance, with kriging and FCNN taking around 39~seconds and 42.6~seconds, respectively. 

\begin{figure}[!htbp]
        \centering
        \includegraphics[width=1\columnwidth]{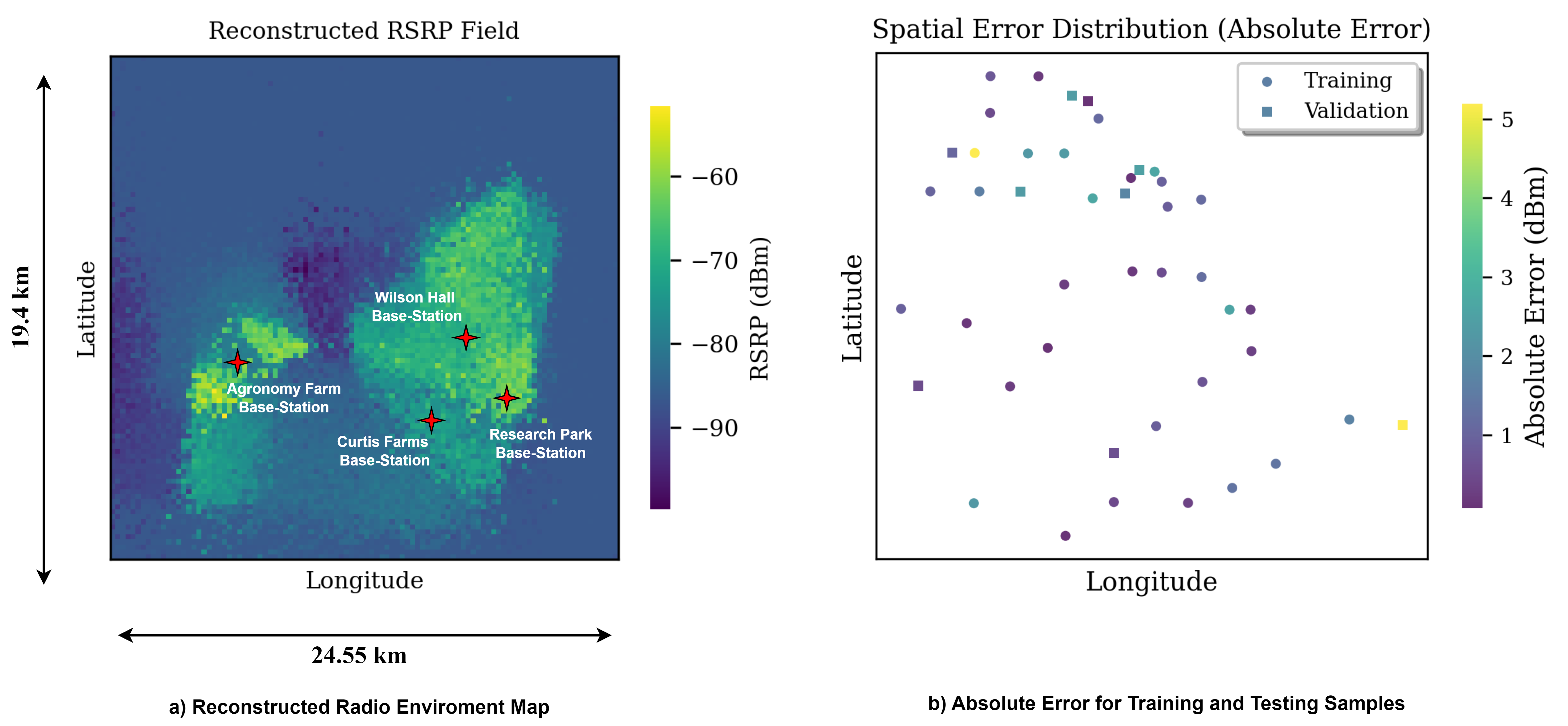}
        \caption{Comparison of Test and Training Error over Regenerated Field 
        %\hz{change b): "Train" to "Training", "Test" to "Testing"}
        }
        \label{fig: Error_comparison}
%        \vspace{-13pt}
\end{figure}

To gain insight into Reveal-MT's modeling accuracy, Fig.~\ref{fig: Error_comparison}(a) 
presents the reconstructed RSSI map for $\lambda = 0.459$. The spatial variation in signal strength is evident, with a distinct high-signal region surrounded by weaker areas.  Such a spatial variability demonstrates ReVeal-MT's ability to effectively capture signal propagation patterns.  The map demonstrates a physically realistic gradient of received power, with clear separation between high-signal regions in close proximity to the transmitter and lower-signal regions farther away or obstructed by terrain. Importantly, the reconstructed field does not exhibit artificial artifacts or over smoothing, which are common in purely statistical interpolation methods such as kriging. Instead, the spatial variations align closely with expected propagation characteristics including the pattern of three distinct radio sectors as expected from the real ARA deployment. This balance between smoothness and heterogeneity illustrates the advantage of embedding PDE-based constraints within the learning framework.

\begin{figure}[!t]
        \centering
        \includegraphics[width=0.7\columnwidth]{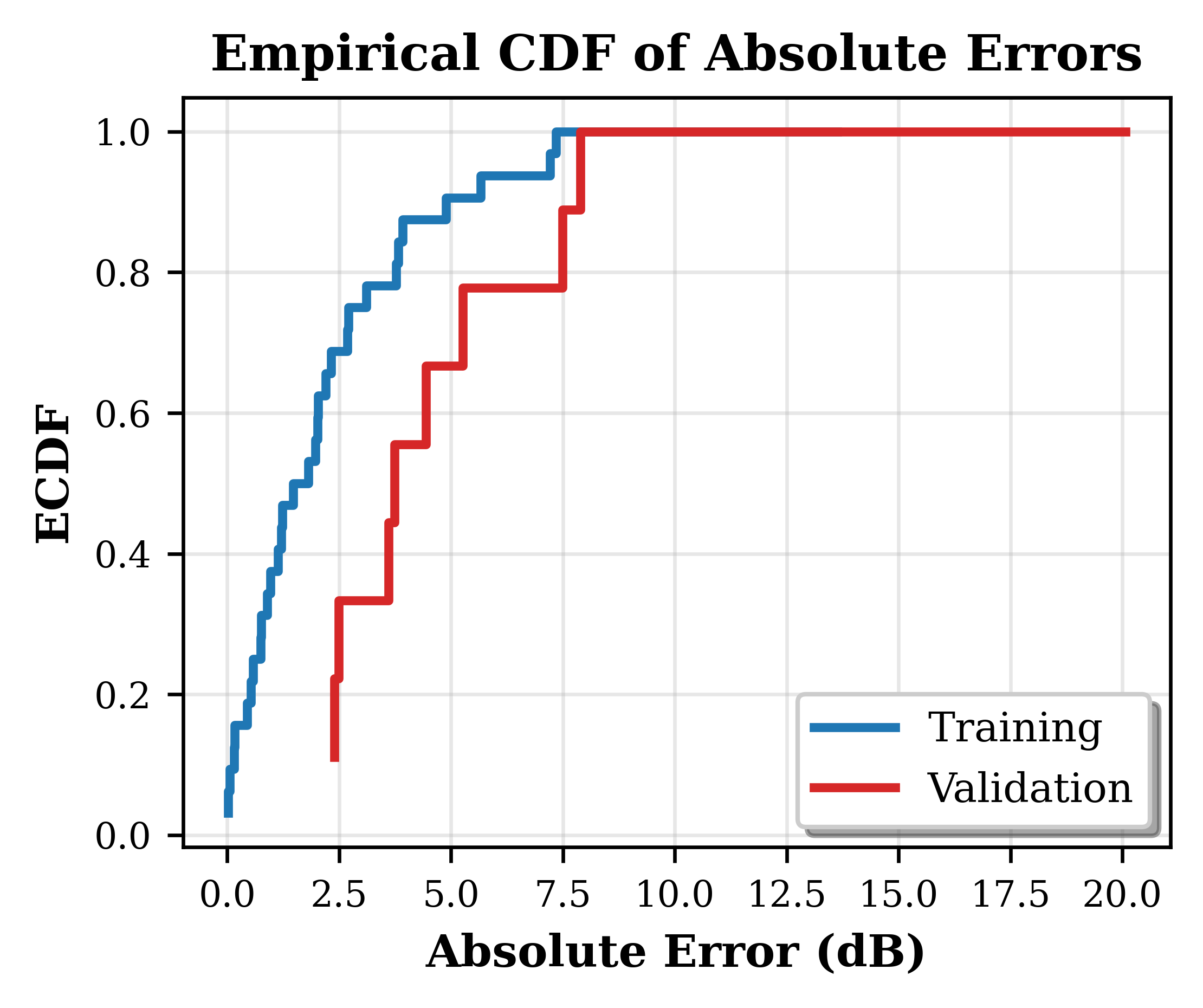}
        \caption{ECDF of Absolute Errors 
        %\hz{1) change ylable to "CDF" 2) remove the legend and mark 0.25 and 0.75 in the y-axis and figure itself.}
        }
        \label{fig: CDF}
%        \vspace{-13pt}
\end{figure}

\begin{comment}

\end{comment}

\begin{figure}[!b]
        \centering
        \includegraphics[width=.8\columnwidth]{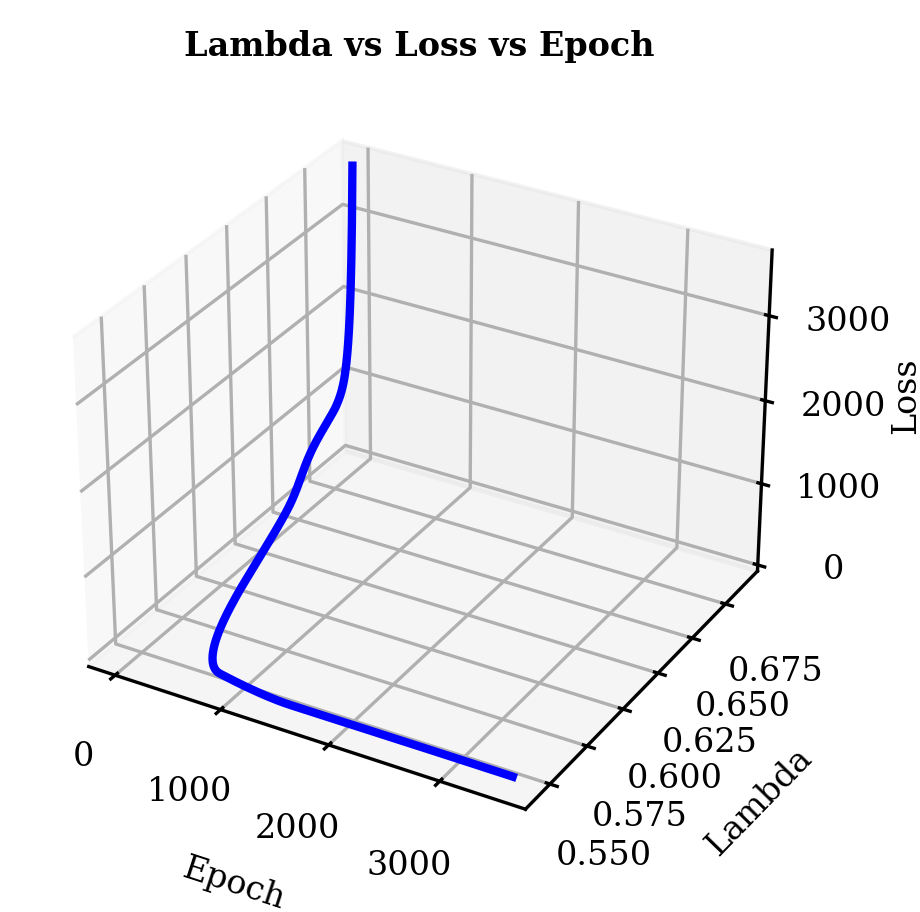}
        \caption{Impact of Varying $\lambda$ on ReVeal-MT Performance }
        \label{fig: lambda_comparison}
%        \vspace{-13pt}
\end{figure}

\begin{figure*}[!t]
    \centering
    \includegraphics[width=0.8\textwidth]{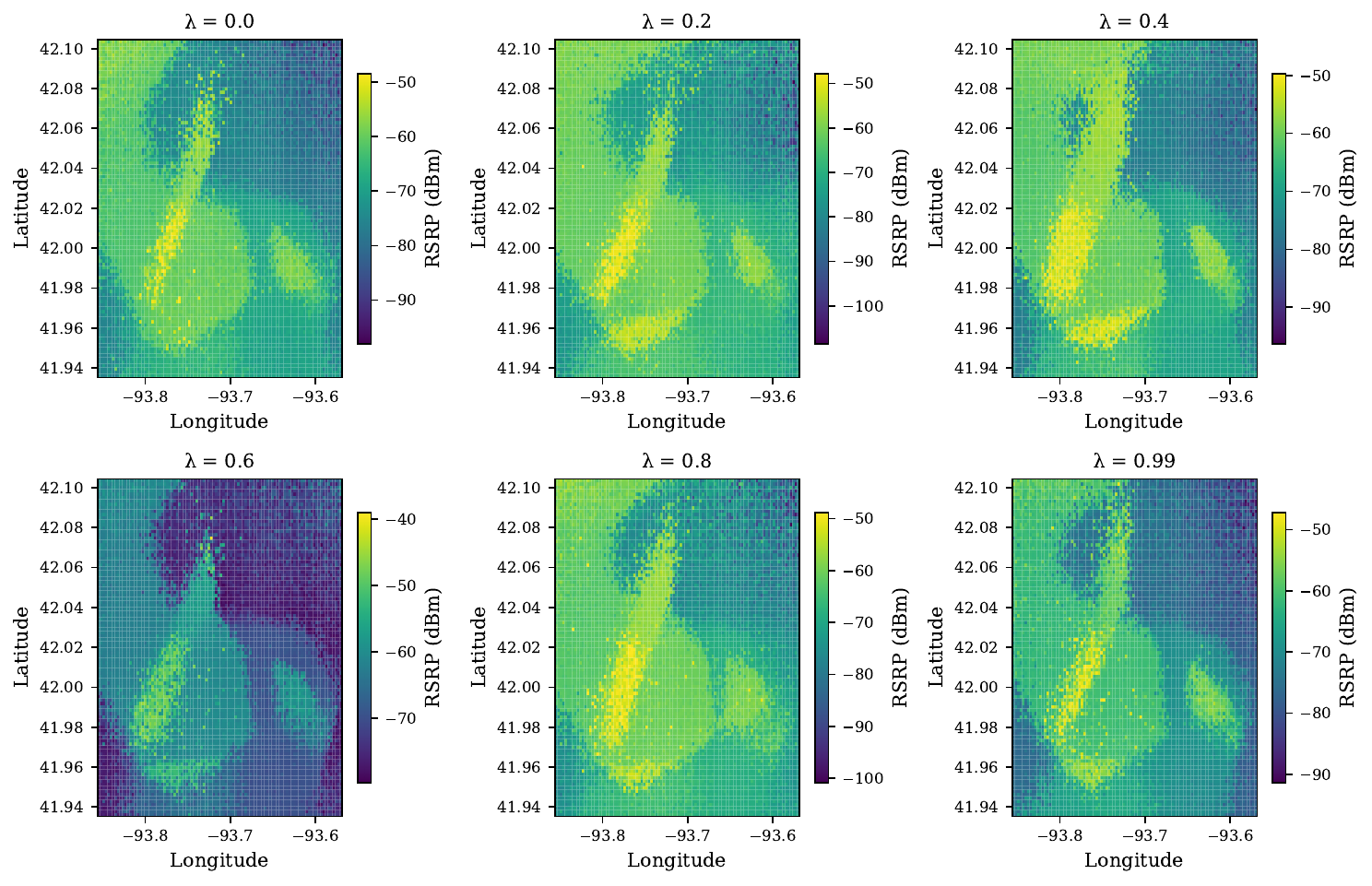}
    \caption{Regenerated Field Under different Lambda. 
    }
    \label{fig:lambda_field}
    % \vspace{-13pt}
\end{figure*}

Fig.~\ref{fig: Error_comparison}~(b)  provides a quantitative view of reconstruction accuracy by mapping the absolute error at both training points (circles) and testing points (squares). Across most of the domain, the error remains below 3 dB, with only a few isolated locations experiencing slightly higher deviations. Crucially, the error distribution shows no systematic bias between training and test points, indicating that ReVeal-MT generalizes well and avoids spatial overfitting. This confirms that the model is not merely memorizing training samples but is effectively interpolating the radio environment in regions without measurements—an essential requirement for spectrum cartography under sparse sensing.

The statistical robustness of these observations is further validated in Fig.~\ref{fig: CDF}, which presents the Empirical Cumulative Distribution Function (ECDF) of absolute errors. The ECDF curve is steep and well-concentrated, with the 25th, 50th, and 75th percentiles at 1.02 dB, 1.31 dB, and 2.39 dB, respectively. This compact distribution of errors demonstrates that the vast majority of predictions lie within a narrow error band, with very few outliers. Such consistency across diverse terrain types, ranging from open farmland to suburban environments emphasizes the model’s ability to capture terrain-induced heterogeneity while maintaining low error margins. The tight error bounds of ReVeal-MT therefore underscore its reliability and robustness for practical deployment in large-scale, real-world spectrum sharing applications.

The effect of varying the weighting parameter $\lambda$ in Eqn.~\eqref{eq:Lp} is depicted in Figs.~\ref{fig: lambda_comparison} and ~\ref{fig:lambda_field}. Fig. ~\ref{fig:lambda_field} visualizes reconstructed fields under different $\lambda$ values, ranging from purely data-driven ($\lambda$ = 0) to physics-dominated ($\lambda$ = 0.99). At lower $\lambda$ values, the model overfits to local sensor observations and fails to generalize globally. At excessively high $\lambda$ values, the model disregards empirical data and drifts away from realistic measurements.

Fig.~\ref{fig: lambda_comparison} presents a three-dimensional view of the training dynamics, showing the joint evolution of $\lambda$, loss, and epochs. The plot clearly demonstrates that ReVeal-MT achieves rapid loss reduction during the initial training phase, particularly when $\lambda$ lies within the range of 0.45–0.55, where the contributions of the data-driven and physics-driven loss terms are effectively balanced. As training progresses, the curve flattens, indicating convergence to a stable solution with consistently low loss values. In contrast, deviations from this balanced $\lambda$ regime lead to sharp increases in loss, emphasizing the sensitivity of the optimization process to the choice of $\lambda$. These results reinforce the findings in Fig. 13, highlighting that an appropriately tuned $\lambda$ is critical to ensuring both stable convergence and high-accuracy spectrum reconstruction in multi-transmitter environments.

\begin{figure}[!b]
        \centering
        \includegraphics[width=1\columnwidth]{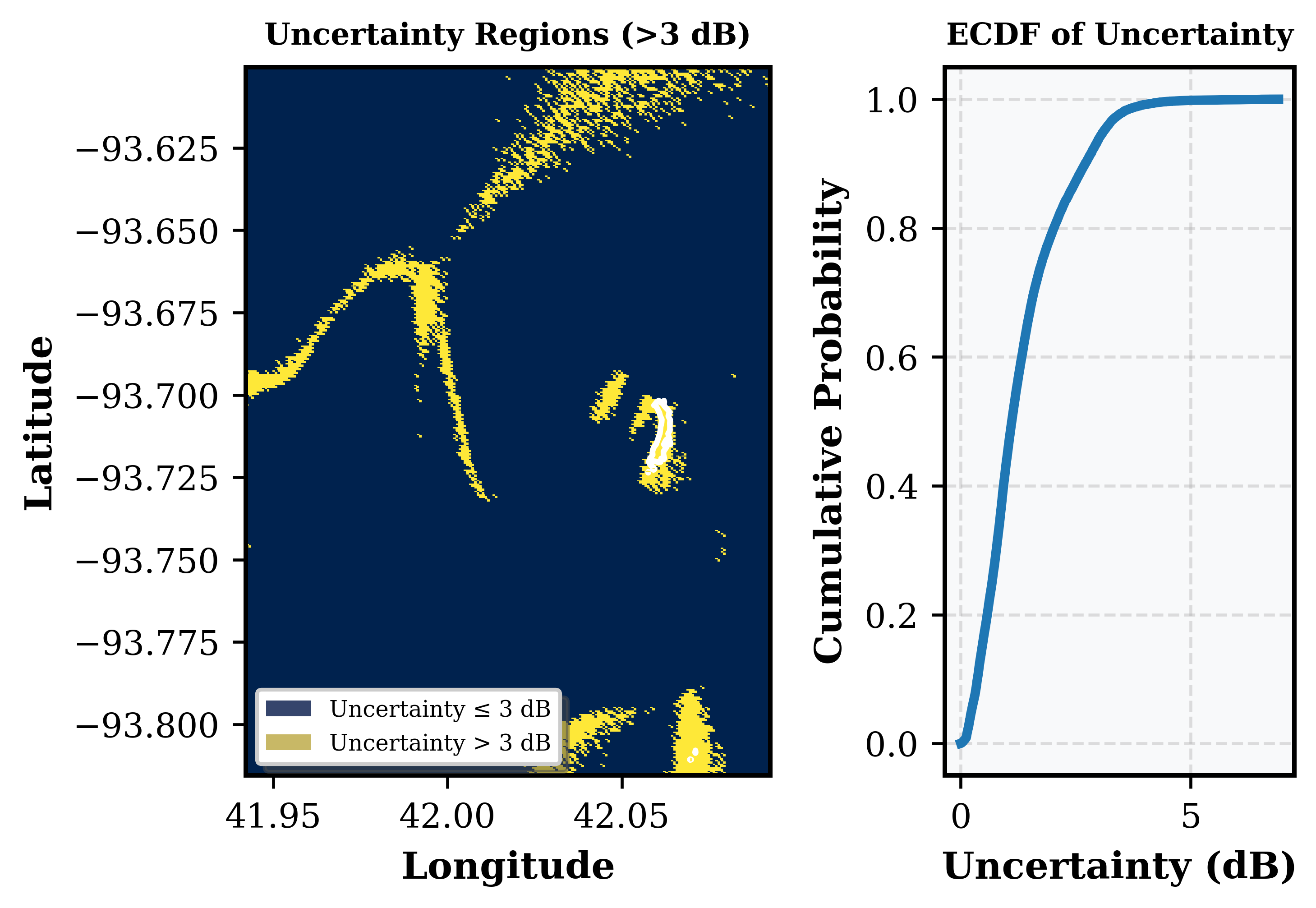}
        \caption{Uncertainty analysis of the reconstructed coverage map. (Left) Spatial regions where model prediction uncertainty exceeds 3 dB, shown in yellow. (Right) Empirical cumulative distribution function (ECDF) of uncertainty across all locations}
        \label{fig: uncertanity_analysis}
%        \vspace{-13pt}
\end{figure}

The predictive uncertainty for the ReVeal-MT model was quantified using Monte Carlo (MC) Dropout, an established approach for estimating epistemic uncertainty in deep learning models. During inference, dropout layers were kept active, and 50 stochastic forward passes were performed for each input; the standard deviation of the resulting predictions was then computed as the uncertainty estimate \cite{Dropout}, \cite{Dropout2}. 

Fig.~\ref{fig: uncertanity_analysis} illustrates the spatial and statistical distribution of uncertainty in the ReVeal-MT path loss model. The analysis identifies specific geographical regions, where the predictive uncertainty exceeds the 3 dB threshold. These areas of high uncertainty, highlight locations where the model's predictions maybe less reliable, potentially due to factors such as sparse training data or complex, unmodeled propagation phenomena. This spatial mapping is crucial for identifying zones where more potential samples must be taken to improve the prediction bounds.

\begin{figure}[!b]
        \centering
        \includegraphics[width=1\columnwidth]{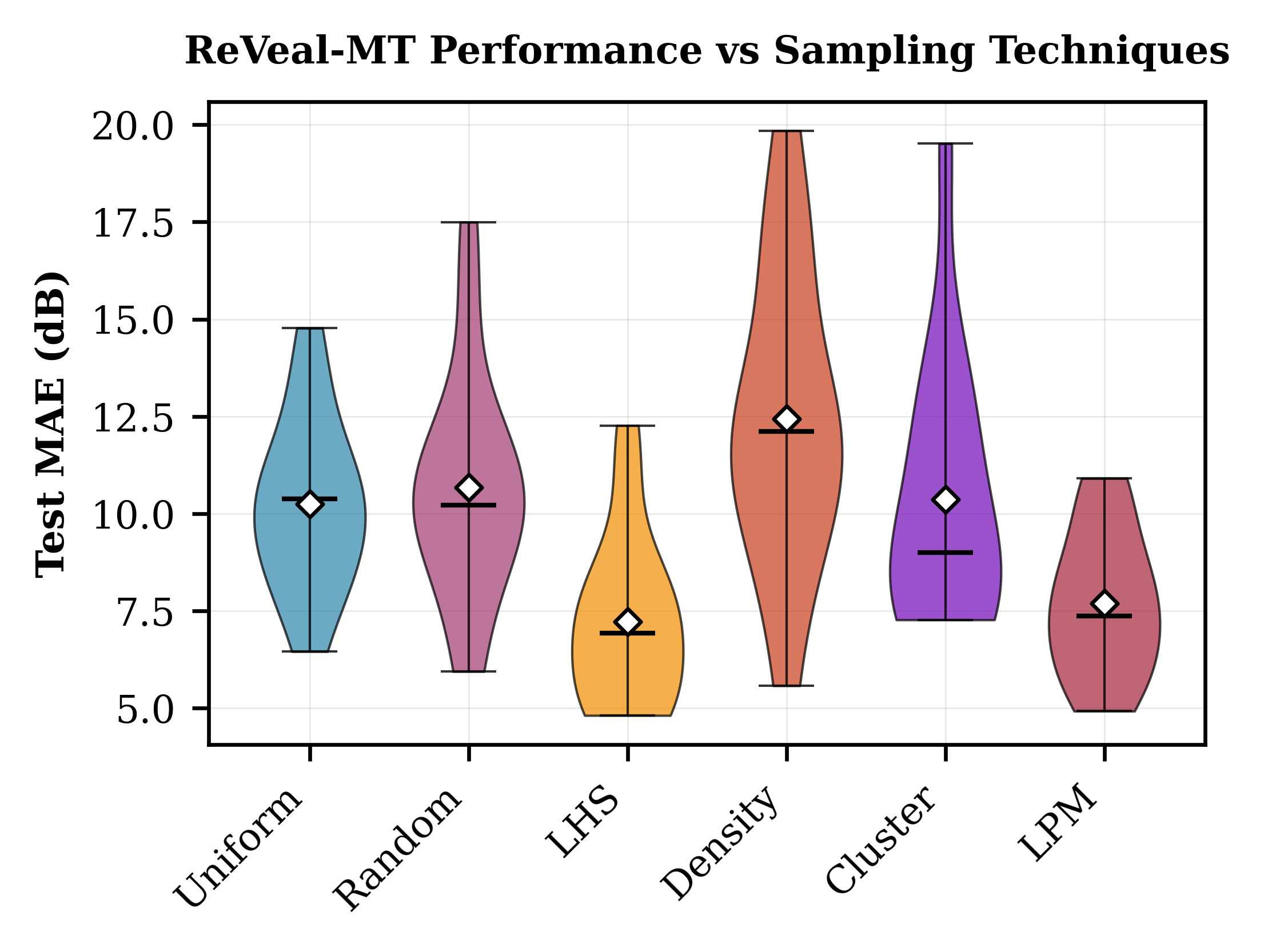}
        \caption{Comparison of test MAE (dB) for different sampling techniques}
        \label{fig: sampling_comparison}
%        \vspace{-13pt}
\end{figure}

Fig.~\ref{fig: sampling_comparison} compares the generalization performance of the ReVeal-MT model, measured by Test Mean Absolute Error (MAE) when trained on datasets constructed using different sampling techniques. The methods evaluated include standard approaches like Uniform and Random sampling, Latin Hypercube Sampling (LHS), as well as more sophisticated strategies such as Density-based, Cluster-based, and Local Pivotal Method (LPM) sampling. The comparative visualization clearly demonstrates that the choice of sampling strategy has a substantial impact on the model's final prediction accuracy, underscoring the importance of active or intelligent data selection beyond simple random acquisition for constructing efficient and accurate radio frequency maps.

\section*{Appendix A}

We now provide the detailed derivation of the governing partial differential equation (PDE) for the received signal strength in the multi-transmitter case.  

The received power at location $(x,y)$ from transmitter $i$ in dB scale is modeled using the standard log-distance path-loss with shadowing.  
\begin{equation}
P_{r,i}^{\mathrm{dB}}(x,y) = P_{T,i}^{\mathrm{dB}} - 10\eta \log_{10}\!\left(\frac{r_i(x,y)}{d_0}\right) + Z_{\sigma,i}(x,y),
\label{eqn:single_tx_dB}
\end{equation}
where
\begin{itemize}
    \item $P_{T,i}^{\mathrm{dB}}$ is the transmit power of transmitter $i$ in dB at $d_0$,
    \item $\eta$ is the path-loss exponent,
    \item $r_i(x,y) = \sqrt{(x-x_i)^2 + (y-y_i)^2}$ is the distance from transmitter $i$,
    \item $d_0$ is the reference distance, and
    \item $Z_{\sigma,i}(x,y)$ is the log-normal shadowing term (Gaussian in dB).
\end{itemize}
Since powers from multiple transmitters superimpose in the \textbf{linear domain}, we first convert Eqn.~\eqref{eqn:single_tx_dB} to linear power.
\begin{equation}
P_{r,i}^{\mathrm{linear}}(x,y) = 10^{P_{r,i}^{\mathrm{dB}}(x,y)/10}.
\label{eqn:dB_to_linear}
\end{equation}
Substituting the expression for $P_{r,i}^{\mathrm{dB}}(x,y)$,
\begin{equation}
P_{r,i}^{\mathrm{linear}}(x,y) = 10^{\tfrac{1}{10}\Big(P_{T,i}^{\mathrm{dB}} - 10\eta \log_{10}\tfrac{r_i(x,y)}{d_0} + Z_{\sigma,i}(x,y)\Big)}.
\label{eqn:linear_single}
\end{equation}
For compactness, define
\begin{equation}
a_i(x,y) \triangleq \tfrac{1}{10}\Big(P_{T,i}^{\mathrm{dB}} - 10\eta \log_{10}\tfrac{r_i(x,y)}{d_0} + Z_{\sigma,i}(x,y)\Big),
\label{eqn:ai_def}
\end{equation}
so that the linear received power reduces to
\begin{equation}
P_{r,i}^{\mathrm{linear}}(x,y) = 10^{a_i(x,y)}.
\label{eqn:linear_ai}
\end{equation}
The total received power from $M$ transmitters becomes
\begin{equation}
S(x,y) \triangleq \sum_{i=1}^M P_{r,i}^{\mathrm{linear}}(x,y) = \sum_{i=1}^M 10^{a_i(x,y)}.
\label{eqn:sum_linear}
\end{equation}
Returning to the dB domain, the aggregate received power becomes 
\begin{equation}
P_{r,\mathrm{tot}}^{\mathrm{dB}}(x,y) = 10 \log_{10} S(x,y) 
= 10 \log_{10}\!\Bigg(\sum_{i=1}^M 10^{a_i(x,y)}\Bigg).
\end{equation}
For subsequent derivatives, define the normalized weight of transmitter $i$ as
\begin{equation}
w_i(x,y) \triangleq \frac{10^{a_i(x,y)}}{\sum_{j=1}^M 10^{a_i(x,y)}}
\end{equation}
These weights represent the \textit{fractional contribution} of each transmitter to the total received power at $(x,y)$. The Laplacian operator in two dimensions is
\begin{equation}
\nabla^2 = \frac{\partial^2}{\partial x^2} + \frac{\partial^2}{\partial y^2}.
\label{eqn:laplacian}
\end{equation}
Using $P_{r,\mathrm{tot}}^{\mathrm{dB}} = \tfrac{10}{\ln 10}\ln S$, we have
\begin{equation}
\nabla^2 P_{r,\mathrm{tot}}^{\mathrm{dB}} 
= \frac{10}{\ln 10} \nabla^2 (\ln S) 
= \frac{10}{\ln 10}\left(\frac{\nabla^2 S}{S} - \frac{\|\nabla S\|^2}{S^2}\right)
\label{eqn:laplnS}
\end{equation}
Since $S = \sum_i 10^{a_i} = \sum_i e^{b a_i}$ with $b=\ln 10$,
\begin{align}
\nabla S &= b \sum_i 10^{a_i}\nabla a_i,\\ 
\nabla^2 S &= b \sum_i 10^{a_i}\nabla^2 a_i 
+ b^2 \sum_i 10^{a_i}\|\nabla a_i\|^2
\label{eqn:gradS}
\end{align}
Substituting these expressions into Eqn.~\eqref{eqn:laplnS} and simplifying using the weights $w_i(x,y)$, the Laplacian of the total received power in dB scale becomes
\begin{align}
\nabla^2 P_{r,\mathrm{tot}}^{\mathrm{dB}} = 10  \sum_i w_i \nabla^2 a_i + 10 \ln& 10  \Big( \sum_i w_i \|\nabla a_i\|^2 \nonumber\\ & - \|\sum_i w_i \nabla a_i\|^2 \Big)
\label{eqn:final1}
\end{align}

\section{Concluding Remarks}
\label{sec:conclusion}

Accurate radio environment mapping is crucial for dynamic spectrum sharing, particularly in complex multi-transmitter scenarios. This paper presented ReVeal-MT, a physics-informed neural network that advances spectrum cartography by integrating a novel multi-transmitter PDE formulation. This approach enables high-fidelity reconstruction of the radio environment from sparse measurements without prior transmitter knowledge. Extensive validation on the ARA wireless living lab demonstrated ReVeal-MT's superior performance, achieving an RMSE of 2.66 dB over 370 km². The model's governing PDE effectively captures the spatial superposition and shadowing effects of multiple transmitters, leading to significant gains over statistical, deterministic, and data-driven benchmarks. ReVeal-MT offers a robust, data-efficient solution for large-scale spectrum management. Future work will explore incorporating temporal dynamics and applying the model to design dynamic spectrum sharing mechanisms.

\section*{Acknowledgment}
{This work is supported in part by the NSF awards 2130889,  2112606, 2212573, 2229654, and 2232461, NIFA award 202167021-33775, and PAWR Industry Consortium. The authors would also like to thank the Microsoft Research team, particularly Ranveer Chandra, Tusher Chakraborty, and Suraj Jog, for intriguing discussions on related topics. %feedback and insightful contributions.
}

\bibliographystyle{IEEEtran}
\bibliography{references}

\begin{thebibliography}{10}
\providecommand{\url}[1]{#1}
\csname url@samestyle\endcsname
\providecommand{\newblock}{\relax}
\providecommand{\bibinfo}[2]{#2}
\providecommand{\BIBentrySTDinterwordspacing}{\spaceskip=0pt\relax}
\providecommand{\BIBentryALTinterwordstretchfactor}{4}
\providecommand{\BIBentryALTinterwordspacing}{\spaceskip=\fontdimen2\font plus
\BIBentryALTinterwordstretchfactor\fontdimen3\font minus \fontdimen4\font\relax}
\providecommand{\BIBforeignlanguage}[2]{{%
\expandafter\ifx\csname l@#1\endcsname\relax
\typeout{** WARNING: IEEEtran.bst: No hyphenation pattern has been}%
\typeout{** loaded for the language `#1'. Using the pattern for}%
\typeout{** the default language instead.}%
\else
\language=\csname l@#1\endcsname
\fi
#2}}
\providecommand{\BIBdecl}{\relax}
\BIBdecl

\bibitem{shahid2024}
\BIBentryALTinterwordspacing
M.~Shahid, K.~Das, T.~U. Islam, C.~Somiah, D.~Qiao, A.~Ahmad, J.~Song, Z.~Zhu, S.~Babu, Y.~Guan, T.~Chakraborty, S.~Jog, R.~Chandra, and H.~Zhang, ``Wireless spectrum in rural farmlands: Status, challenges and opportunities,'' Jul. 2024. [Online]. Available: \url{https://arxiv.org/abs/2407.04561}
\BIBentrySTDinterwordspacing

\bibitem{DB_Critique}
\BIBentryALTinterwordspacing
R.~Ramjee, S.~Roy, and K.~Chintalapudi, ``A critique of {FCC}'s {TV} white space regulations,'' vol.~20, no.~1, Aug. 2016. [Online]. Available: \url{https://doi.org/10.1145/2972413.2972421}
\BIBentrySTDinterwordspacing

\bibitem{DB_Implementation}
M.~Ante, J.~Molina, E.~Trinidad, and L.~Materum, ``A survey and comparison of {TV} white space implementations in {Japan, the Philippines, Singapore, the united kingdom, and the United States},'' \emph{International Journal of Advanced Technology and Engineering Exploration}, vol.~8, pp. 780--796, Jul. 2021.

\bibitem{Bhattarai2018}
\BIBentryALTinterwordspacing
S.~Bhattarai, ``Spectrum efficiency and security in dynamic spectrum sharing,'' Ph.D. dissertation, Virginia Tech, Apr. 2018, doctoral Dissertation. [Online]. Available: \url{http://hdl.handle.net/10919/82872}
\BIBentrySTDinterwordspacing

\bibitem{fcc2020}
\BIBentryALTinterwordspacing
{Federal Communications Commission}, ``Facilitating shared use in the 3.1--3.55 {GHz} band,'' Jan. 2020, federal Register, Vol. 85, No. 14, Pages 3888-3903. [Online]. Available: \url{https://www.federalregister.gov/documents/2020/01/22/2020-00535/facilitating-shared-use-in-the-31-355-ghz-band}
\BIBentrySTDinterwordspacing

\bibitem{Spectrum_carography}
\BIBentryALTinterwordspacing
Y.~S. Reddy, A.~Kumar, O.~J. Pandey, and L.~R. Cenkeramaddi, ``Spectrum cartography techniques, challenges, opportunities, and applications: A survey,'' \emph{Pervasive Mob. Comput.}, vol.~79, no.~C, Jan. 2022. [Online]. Available: \url{https://doi.org/10.1016/j.pmcj.2021.101511}
\BIBentrySTDinterwordspacing

\bibitem{cartography_techniques}
\BIBentryALTinterwordspacing
Y.~Teganya, D.~Romero, L.~M.~L. Ramos, and B.~Beferull-Lozano, ``Location-free spectrum cartography,'' \emph{Trans. Sig. Proc.}, vol.~67, no.~15, p. 4013–4026, Aug. 2019. [Online]. Available: \url{https://doi.org/10.1109/TSP.2019.2923151}
\BIBentrySTDinterwordspacing

\bibitem{Subash}
\BIBentryALTinterwordspacing
S.~Timilsina, S.~Shrestha, and X.~Fu, ``Quantized radio map estimation using tensor and deep generative models,'' \emph{Trans. Sig. Proc.}, vol.~72, pp. 173–--189, Nov. 2023. [Online]. Available: \url{https://doi.org/10.1109/TSP.2023.3336179}
\BIBentrySTDinterwordspacing

\bibitem{SC_survey}
D.~Romero and S.-J. Kim, ``Radio map estimation: A data-driven approach to spectrum cartography,'' \emph{IEEE Signal Processing Magazine}, vol.~39, no.~6, pp. 53--72, Nov. 2022.

\bibitem{NTIA_spectrum_cartography}
\BIBentryALTinterwordspacing
C.~R. Dietlein, ``Wide-area spectrum cartography,'' National Telecommunications and Information Administration, Institute for Telecommunication Sciences, Boulder, CO, USA, Tech. Rep., accessed: 2024-11-21. [Online]. Available: \url{https://its.ntia.gov/}
\BIBentrySTDinterwordspacing

\bibitem{ReVeal}
M.~Shahid, K.~Das, H.~Ushaq, H.~Zhang, J.~Song, D.~Qiao, S.~Babu, Y.~Guan, Z.~Zhu, and A.~Ahmad, ``Reveal: A physics-informed neural network for high-fidelity radio environment mapping,'' in \emph{2025 IEEE International Symposium on Dynamic Spectrum Access Networks (DySPAN)}, 2025, pp. 1--10.

\bibitem{GNN_Multi_Source}
\BIBentryALTinterwordspacing
X.~Wen, S.~Fang, and Y.~Fan, ``Reconstruction of radio environment map based on multi-source domain adaptive of graph neural network for regression,'' \emph{Sensors}, vol.~24, no.~8, Jun. 2024. [Online]. Available: \url{https://www.mdpi.com/1424-8220/24/8/2523}
\BIBentrySTDinterwordspacing

\bibitem{Block_tensor_decomposition}
H.~Sun and J.~Chen, ``Integrated interpolation and block-term tensor decomposition for spectrum map construction,'' \emph{IEEE Transactions on Signal Processing}, vol.~72, pp. 3896--3911, Aug. 2024.

\bibitem{ARA_design_implementation}
\BIBentryALTinterwordspacing
T.~U. Islam, J.~O. Boateng, M.~Nadim, G.~Zu, M.~Shahid, X.~Li, T.~Zhang, S.~Reddy, W.~Xu, A.~Atalar, V.~Lee, Y.-F. Chen, E.~Gosling, E.~Permatasari, C.~Somiah, Z.~Meng, S.~Babu, M.~Soliman, A.~Hussain, D.~Qiao, M.~Zheng, O.~Boyraz, Y.~Guan, A.~Arora, M.~Selim, A.~Ahmad, M.~B. Cohen, M.~Luby, R.~Chandra, J.~Gross, and H.~Zhang, ``Design and implementation of {ARA} wireless living lab for rural broadband and applications,'' Aug. 2024. [Online]. Available: \url{https://arxiv.org/abs/2408.00913}
\BIBentrySTDinterwordspacing

\bibitem{Channel_measurement_survey}
\BIBentryALTinterwordspacing
J.~Zhang, J.~Lin, P.~Tang, Y.~Zhang, H.~Xu, T.~Gao, H.~Miao, Z.~Chai, Z.~Zhou, Y.~Li, H.~Gong, Y.~Liu, Z.~Yuan, X.~Liu, L.~Tian, S.~Yang, L.~Xia, G.~Liu, and P.~Zhang, ``Channel measurement, modeling, and simulation for {6G}: A survey and tutorial,'' \emph{arXiv preprint arXiv:2305.16616}, Oct. 2023. [Online]. Available: \url{https://arxiv.org/abs/2305.16616}
\BIBentrySTDinterwordspacing

\bibitem{Tataria2020}
\BIBentryALTinterwordspacing
H.~Tataria, K.~Haneda, A.~F. Molisch, M.~Shafi, and F.~Tufvesson, ``Standardization of propagation models for terrestrial cellular systems: A historical perspective,'' \emph{International Journal of Wireless Information Networks}, vol.~27, pp. 340--364, Mar. 2020. [Online]. Available: \url{https://link.springer.com/article/10.1007/s10776-020-00500-9}
\BIBentrySTDinterwordspacing

\bibitem{ray_tracing_based_model}
J.-H. Lee and A.~F. Molisch, ``A scalable and generalizable pathloss map prediction,'' \emph{IEEE Transactions on Wireless Communications}, vol.~23, no.~11, pp. 17\,793--17\,806, Nov. 2024.

\bibitem{Nueral_Ray_Tracing}
\BIBentryALTinterwordspacing
T.~Orekondy, P.~Kumar, S.~Kadambi, H.~Ye, J.~Soriaga, and A.~Behboodi, ``{WiNeRT}: Towards neural ray tracing for wireless channel modelling and differentiable simulations,'' in \emph{Proceedings of the 11th International Conference on Learning Representations (ICLR 2023)}, Jul. 2023. [Online]. Available: \url{https://openreview.net/forum?id=tPKKXeW33YU}
\BIBentrySTDinterwordspacing

\bibitem{RDZ}
M.~Zheleva, C.~R. Anderson, M.~Aksoy, J.~T. Johnson, H.~Affinnih, and C.~G. DePree, ``Radio dynamic zones: Motivations, challenges, and opportunities to catalyze spectrum coexistence,'' \emph{IEEE Communications Magazine}, vol.~61, no.~6, pp. 156--162, Jan. 2023.

\bibitem{Kriging}
P.~Maiti and D.~Mitra, ``Complexity reduction of ordinary {Kriging} algorithm for {3D REM} design,'' \emph{Physical Communication}, vol.~55, p. 101912, Oct. 2022.

\bibitem{DeepREM}
A.~Chaves-Villota and C.~A. Viteri-Mera, ``{DeepREM}: Deep-learning-based radio environment map estimation from sparse measurements,'' \emph{IEEE Access}, vol.~11, pp. 48\,697--48\,714, May. 2023.

\bibitem{ProSpire}
S.~Sarkar, D.~Guo, and D.~Cabric, ``{ProSpire}: Proactive spatial prediction of radio environment using deep learning,'' in \emph{2023 20th Annual IEEE International Conference on Sensing, Communication, and Networking (SECON)}, Dec. 2023, pp. 177--185.

\bibitem{CNN}
H.~Cheng, H.~Lee, and S.~Ma, ``{CNN}-based indoor path loss modeling with reconstruction of input images,'' in \emph{2018 International Conference on Information and Communication Technology Convergence (ICTC)}, Nov. 2018, pp. 605--610.

\bibitem{U_NET}
M.~Mallik, S.~Kharbech, T.~Mazloum, S.~Wang, J.~Wiart, D.~P. Gaillot, and L.~Clavier, ``{EME-Net}: A {U-net-based} indoor {EMF} exposure map reconstruction method,'' in \emph{2022 16th European Conference on Antennas and Propagation (EuCAP)}, Jun. 2022, pp. 1--5.

\bibitem{GAN}
\BIBentryALTinterwordspacing
X.~Han, L.~Xue, F.~Shao, and Y.~Xu, ``A power spectrum maps estimation algorithm based on generative adversarial networks for underlay cognitive radio networks,'' \emph{Sensors}, vol.~20, no.~1, Aug. 2020. [Online]. Available: \url{https://www.mdpi.com/1424-8220/20/1/311}
\BIBentrySTDinterwordspacing

\bibitem{PINN_to_PIKAN}
\BIBentryALTinterwordspacing
J.~D. Toscano, V.~Oommen, A.~J. Varghese, Z.~Zou, N.~A. Daryakenari, C.~Wu, and G.~E. Karniadakis, ``From {PINNs to PIKANs}: Recent advances in physics-informed machine learning,'' \emph{arXiv preprint arXiv:2410.13228}, Oct. 2024. [Online]. Available: \url{https://arxiv.org/abs/2410.13228}
\BIBentrySTDinterwordspacing

\bibitem{Possion_FEM}
\BIBentryALTinterwordspacing
D.~E. Johnson, ``Numerical solutions to {Poisson} equations using the finite-difference method,'' \emph{IEEE Antennas and Propagation Magazine}, vol.~56, no.~6, pp. 158--162, Jan. 2014. [Online]. Available: \url{https://ieeexplore.ieee.org/document/6931698}
\BIBentrySTDinterwordspacing

\bibitem{PINN_RAISSI}
\BIBentryALTinterwordspacing
M.~Raissi, P.~Perdikaris, and G.~Karniadakis, ``Physics-informed neural networks: A deep learning framework for solving forward and inverse problems involving nonlinear partial differential equations,'' \emph{Journal of Computational Physics}, vol. 378, pp. 686--707, Feb. 2019. [Online]. Available: \url{https://www.sciencedirect.com/science/article/pii/S0021999118307125}
\BIBentrySTDinterwordspacing

\bibitem{PINN_poisson}
\BIBentryALTinterwordspacing
Y.~Zhang, J.~Li, Y.~Wang, X.~Zhang, Y.~Zhang, and J.~Zhang, ``Physics informed neural network for charged particles surrounded by conductive walls,'' \emph{Scientific Reports}, vol.~13, 2023. [Online]. Available: \url{https://www.nature.com/articles/s41598-023-40477-y}
\BIBentrySTDinterwordspacing

\bibitem{PINN_wireless_1}
S.~{Wagle}, A.~{Malhotra}, S.~{Hamidi-Rad}, A.~{Sant}, D.~J. {Love}, and C.~G. {Brinton}, ``{Physics-Informed Generative Approaches for Wireless Channel Modeling},'' \emph{arXiv e-prints}, p. arXiv:2503.05988, Mar. 2025.

\bibitem{PINN_Wireless_2}
F.~Jiang, T.~Li, X.~Lv, H.~Rui, and D.~Jin, ``Physics-informed neural networks for path loss estimation by solving electromagnetic integral equations,'' \emph{IEEE Transactions on Wireless Communications}, vol.~23, no.~10, pp. 15\,380--15\,393, 2024.

\bibitem{PINN_wireless_3}
P.~Mithillesh~Kumar and M.~Supriya, ``Attenuation modeling using physics guided deep reinforcement learning: A channel estimation use case,'' \emph{IEEE Open Journal of the Communications Society}, vol.~6, pp. 3696--3709, 2025.

\bibitem{OJAP_PINN}
O.~Noakoasteen, S.~Wang, Z.~Peng, and C.~Christodoulou, ``Physics-informed deep neural networks for transient electromagnetic analysis,'' \emph{IEEE Open Journal of Antennas and Propagation}, vol.~1, pp. 404--412, 2020.

\bibitem{PINN_maths}
B.~E. Mokhtari, C.~Chauviere, and P.~Bonnet, ``On the importance of the mathematical formulation to get pinns working,'' \emph{IEEE Transactions on Electromagnetic Compatibility}, vol.~66, no.~6, pp. 2142--2149, 2024.

\bibitem{Autotune}
P.~Koch, O.~Golovidov, S.~Gardner, B.~Wujek, J.~Griffin, and Y.~Xu, ``Autotune: A derivative-free optimization framework for hyperparameter tuning,'' Apr. 2018.

\bibitem{SMAC}
\BIBentryALTinterwordspacing
F.~Hutter, H.~H. Hoos, and K.~Leyton-Brown, ``Sequential model-based optimization for general algorithm configuration,'' in \emph{Proceedings of the 5th International Conference on Learning and Intelligent Optimization}, ser. LION'05.\hskip 1em plus 0.5em minus 0.4em\relax Berlin, Heidelberg: Springer-Verlag, Jan. 2011, p. 507–523. [Online]. Available: \url{https://doi.org/10.1007/978-3-642-25566-3_40}
\BIBentrySTDinterwordspacing

\bibitem{optuna}
T.~Akiba, S.~Sano, T.~Yanase, T.~Ohta, and M.~Koyama, ``{O}ptuna: A next-generation hyperparameter optimization framework,'' in \emph{The 25th ACM SIGKDD International Conference on Knowledge Discovery \& Data Mining}, Mar. 2019, pp. 2623--2631.

\bibitem{spatial_sampling}
\BIBentryALTinterwordspacing
A.~Ivanov, K.~Tonchev, V.~Poulkov, A.~Manolova, and A.~Vlahov, ``Limited sampling spatial interpolation evaluation for {3D} radio environment mapping,'' \emph{Sensors}, vol.~23, no.~22, Sep. 2023. [Online]. Available: \url{https://www.mdpi.com/1424-8220/23/22/9110}
\BIBentrySTDinterwordspacing

\bibitem{grafstra2012spatially}
A.~Grafstr{\~A}, N.~L. Lundstr{\~A}, L.~Schelin \emph{et~al.}, ``Spatially balanced sampling through the pivotal method,'' \emph{Biometrics}, vol.~68, Jun 2012.

\bibitem{3gpp_ts_38.901}
\BIBentryALTinterwordspacing
3GPP, ``{3GPP TS 38.901}---{Study} on channel model for frequencies from {0.5 to 100 GHz},'' Mar. 2021, accessed: 2024-12-13. [Online]. Available: \url{https://portal.3gpp.org/desktopmodules/Specifications/SpecificationDetails.aspx?specificationId=3173}
\BIBentrySTDinterwordspacing

\bibitem{etsi_tr_138901}
\BIBentryALTinterwordspacing
ETSI, ``{ETSI TR 138 901 V14.0.0}--- technical report: Study on channel model for frequencies from {0.5 to 100 GHz},'' Apr. 2020, accessed: 2024-12-13. [Online]. Available: \url{https://www.etsi.org/deliver/etsi_tr/138900_138999/138901/14.00.00_60/tr_138901v140000p.pdf}
\BIBentrySTDinterwordspacing

\bibitem{sionna}
J.~Hoydis, S.~Cammerer, F.~{Ait Aoudia}, A.~Vem, N.~Binder, G.~Marcus, and A.~Keller, ``Sionna: An open-source library for next-generation physical layer research,'' \emph{arXiv preprint}, Mar. 2023.

\bibitem{MathWorks_RayTracing}
{The MathWorks, Inc.}, ``Raytracing ‒ ray tracing propagation model ‒ matlab,'' \url{https://www.mathworks.com/help/comm/ref/rfprop.raytracing.html}, 2025, accessed: 2025-09-22.

\bibitem{Dropout}
\BIBentryALTinterwordspacing
Y.~Gal and Z.~Ghahramani, ``Dropout as a bayesian approximation: Representing model uncertainty in deep learning,'' in \emph{Proceedings of The 33rd International Conference on Machine Learning}, ser. Proceedings of Machine Learning Research, M.~F. Balcan and K.~Q. Weinberger, Eds., vol.~48.\hskip 1em plus 0.5em minus 0.4em\relax New York, New York, USA: PMLR, 20--22 Jun 2016, pp. 1050--1059. [Online]. Available: \url{https://proceedings.mlr.press/v48/gal16.html}
\BIBentrySTDinterwordspacing

\bibitem{Dropout2}
\BIBentryALTinterwordspacing
S.~Alavala and S.~Gorthi, ``Certainty in uncertainty! an improved gi bleeding detection pipeline with uncertainty estimation.''\hskip 1em plus 0.5em minus 0.4em\relax New York, NY, USA: Association for Computing Machinery, 2025. [Online]. Available: \url{https://doi.org/10.1145/3702250.3702271}
\BIBentrySTDinterwordspacing

\end{thebibliography}

\end{document}